\documentclass[aps,superscriptaddress,showpacs,nofootinbib,eqsecnum]{revtex4}
 
\usepackage{epsfig} 
\input{epsf}
 
\begin{document}

\title{Emergence of Tricritical Point and Liquid-Gas Phase in the Massless 2+1
  Dimensional Gross-Neveu Model}

\author{Jean-Lo\"{\i}c Kneur} \email{kneur@lpta.univ-montp2.fr}
\affiliation{Laboratoire de Physique Th\'{e}orique et Astroparticules - CNRS -
  UMR 5207, Universit\'{e} Montpellier II, France}

\author{Marcus Benghi Pinto} \email{marcus@fsc.ufsc.br}
\affiliation{Departamento de F\'{\i}sica, Universidade Federal de Santa
  Catarina, 88040-900 Florian\'{o}polis, Santa Catarina, Brazil}

\author{Rudnei O. Ramos} \email{rudnei@uerj.br} \affiliation{Departamento de
  F\'{\i}sica Te\'orica, Universidade do Estado do Rio de Janeiro, 20550-013
  Rio de Janeiro, RJ, Brazil}

\author{Ederson Staudt} \email{ederson@fsc.ufsc.br} \affiliation{Departamento
  de F\'{\i}sica, Universidade Federal de Santa Catarina, 88040-900
  Florian\'{o}polis, Santa Catarina, Brazil}

\begin{abstract}
  
  A complete thermodynamical analysis of the $2+1$ dimensional massless
  Gross-Neveu model is performed using the optimized perturbation theory.
  This is a non-perturbative method that allows us to go beyond the known
  large-$N$ results already at lowest order.  Our results, for a finite number
  of fermion species, $N$, show the existence of a tricritical point in the
  temperature and chemical potential phase diagram for discrete chiral phase
  transition allowing us to precisely to locate it. By studying the phase
  diagram in the pressure and inverse density plane, we also show the
  existence of a liquid-gas phase, which, so far, was unknown to exist in this
  model. {}Finally, we also derive $N$ dependent analytical expressions for
  the order parameter, critical temperature and critical chemical potential.

\end{abstract}

\pacs{11.10.Wx, 12.38.Cy, 11.15.Tk}

\maketitle

\section{Introduction}

The Gross-Neveu (GN) model \cite{gn} has been extensively used as a prototype
for quantum chromodynamics (QCD) and related issues. This is due to the fact
that both models share some common features such as asymptotic freedom and
chiral symmetry breaking (CSB).  {}For this reason the GN model is useful as a
toy model to test different techniques that can be ultimately used to tackle
problems related to QCD phase transitions. At the same time, in the condensed
matter physics domain, the two (1+1) dimensional GN model (GN2d) has been
associated to polymers \cite{gnpolymers} including unidimensional molecules
such as polyacetylene \cite{muc1n}, while the three (2+1) dimensional version
(GN3d) has been related to planar superconductors \cite{rose}. The
applications concerning phase transitions within the GN model are commonly
carried out in the finite temperature and/or finite density domain where
non-perturbative techniques must be employed. In these applications the most
commonly used analytical non-perturbative technique is the $1/N$ expansion
\cite{coleman} where, for the GN model, $N$ represents the number of fermionic
species (see Ref. \cite{largeNreview} for a recent review). In general, this
expansion is considered only at the leading order in what is known as the
large-$N$ approximation.

In the large-$N$ approximation there are fundamental differences between the
GN2d and GN3d models that are worth recalling.  In dimensions $d=2$ one
observes chiral symmetry restoration (CSR) occurring via a phase transition of
the second kind for high temperature ($T$) and small chemical potential
($\mu$) values, while for low $T$ and high $\mu$ the transition is of the
first kind \cite{wolff}. One finds a tricritical point in the $T-\mu$ plane
separating the second order transition line from the first order one, while
metastable lines accompany the first order transition \cite{wolff,italianos}.
In the $P-1/\rho$ plane ($P$ being the pressure and $\rho$ the density) one
finds a phase diagram similar to the one generated by a Van der Waals liquid
so that the CSB region corresponds to the ``gas'' phase while the CSR region
corresponds to the ``liquid'' phase. It then follows that the first order
transition allows for the appearance of a (mixed) liquid-gas phase.  The
chiral (dynamical) symmetry breaking happening in the massless GN2d with
discrete symmetry at finite temperature, however, must be seen with care,
since it is an artifact of the large-$N$ approximation.  This is a consequence
of well-known no-go theorems concerning that there should be no discrete
symmetry breaking in one-space dimension \cite{landau}.  In this case the
system's vacuum manifold allows for the appearance of kink--anti-kink
configurations that are unsuppressed at any finite temperature \cite{kink}.
The system becomes segmented into regions of alternating signs of the order
parameter whose net average value becomes zero.  At leading order, the $1/N$
approximation misses this effect because the energy per kink goes to infinity
as $N \to \infty$, while the contribution from the kinks has the form
$e^{-N}$.  The large-$N$ results for the GN model in 2+1 dimensions are rather
different \cite{rose}.  {}First of all, as far a discrete chiral symmetry is
concerned, the no-go theorem of one-space dimensions no longer applies (though
now, in two-space dimensions, the no-go theorem applies to the non-existence
of a continuous broken symmetry at any finite temperature). The GN3d phase
diagram produced by the large-$N$ approximation shows that the CSB/CSR
transition is of the second kind everywhere except at $T=0$, where it happens
to be of the first kind.  Within this approximation, there are no tricritical
points lying in the $T-\mu$ plane and no liquid-gas phase.  The model behaves
more like a planar superconductor with the transition CSB/CSR happening as a
superconducting/normal one \cite{rose}.  Later, Kogut and Strouthos have used
lattice Monte Carlo simulations to study the GN3d at finite $N$ \cite
{kogutmc}. They predicted that a tricritical point should exist at low finite
values of $T$, but within the numerical precision of their simulations, they
were unable to give its exact location (in earlier lattice simulations
\cite{hands} used to obtain the phase diagram for the GN3d at finite $N$, some
evidence for a tricritical point was also pointed out).  No other attempts or
approximations were able to improve on this situation. As far we are aware of,
no evidence has been given so far for a possible chiral ``liquid-gas" kind of
phase.

In an attempt to go beyond the simple large-$N$ approximation and in such a
way that temperature and chemical potential effects could be considered in a
simple approximation method, three of the present authors \cite{prdgn2} have
recently considered the GN2d in the linear $\delta$ expansion method (LDE),
also known as the optimized perturbation (OPT).  That study allowed for the
inclusion of the first non trivial finite $N$ corrections to the complete
phase diagram of the GN2d model. The main results of Ref.  \cite{prdgn2} are
the derivation of analytic non-perturbative expressions, containing finite $N$
corrections, scalar field expectation value ,critical temperature $T_c$ (at
$\mu=0$), critical chemical potential $\mu_c$ (at $T=0$), as well as for the
tricritical point (at $T \ne 0$ and $\mu \ne 0$). In the phase diagram, the
predicted CSB region is reduced for finite values of $N$. The OPT expression
for $T_c$ predicts values that are lower than the ones predicted by the
large-$N$ approximation which, in the light of the
Coleman-Mermin-Wagner-Landau theorem \cite{landau}, can be viewed as an
indication of convergence.

Our recent success in treating the GN2d \cite{prdgn2} and the previous lattice
Monte Carlo results on the GN3d, concerning the eventual existence of a
tricritical point in the GN3d model \cite{hands, kogutmc}, gave us the
motivation to investigate the GN3d using the OPT method to fully study its
thermodynamics in order to confirm, in an analytical way, the existence of a
tricritical point.  The OPT method is known for exactly reproducing the
large-$N$ result for the effective potential (or free energy) already at the
first non trivial order \cite{npb,prdgn2}. The {\it perturbative } computation
of higher orders brings finite $N$ corrections and non-perturbative results
are generated upon using a variational criterion. One advantage is that at any
perturbative order one has complete control over the contributions, while the
eventual technical difficulties are like the ones one should encounter in a
traditional perturbative computation.  The convergence properties of the OPT
in critical problems associated to Bose-Einstein condensates have been proved
\cite {new,braaten}. It is worth mentioning that some of the most accurate
numerical results regarding the critical temperature for weakly interacting
homogeneous Bose gases have also been obtained with this method \cite{knp}.
Concerning the GN3d model the results obtained in the present work include
analytical equations for both $T_c$ (at $\mu=0$) and $\mu_c$ (at $T=0$) as
well as for the scalar field vacuum expectation value (vev) ${\bar \sigma}_c$
with finite $N$ corrections. Contrary to the GN2d case, these values appear to
be higher than the predicted large-$N$ values.  One of our most important
results concerns the location of a tricritical point at finite values of $T$
and $\mu$. Being able to specialize to {\it any} value of $N$ we choose $N=3$,
which is the relevant value for QCD.  In a preliminary work \cite{letter}, we
already have shown that the unstable region in the $T-\mu$ plane, which
corresponds to the region inside the metastable lines that accompany the first
order transition line, is rather small, thus explaining the difficulty in
observing and locating the tricritical point in previous works. Here we extend
that work and, from the (Landau) free energy, or effective potential, we
obtain other relevant thermodynamical quantities such as the thermodynamical
potential, pressure, density, etc. This allows us to obtain the phase diagram
in the intuitively more accessible $P-1/\rho$ plane and that shows how
important are the finite $N$ corrections in the GN3d.  In fact, these
corrections produce a phase diagram that is like a Van der Waals liquid and
{\it contrary} to the large-$N$ predictions, we show that the model can
display a mixed ``liquid-gas" phase, which was previously unknown to exist.

This work is organized as follows. In the next section we present the GN
model.  In Sec. III we present the OPT method and the interpolated GN model,
evaluating the effective potential in this non-perturbative scheme. We show
that, already at leading order, our results go beyond the known large-$N$
results.  In Sec. IV we present the optimized results obtained from the
effective potential at finite temperature and chemical potential.  The
dynamically generated fermion mass that we here associate with the
auxiliar scalar field vacuum expectation value, 
the critical temperature and critical
chemical potential for chiral symmetry restoration are evaluated and explicit
analytical expressions for these quantities are obtained.  In this same
section we also discuss the complete phase diagram for the GN model in the $T$
and $\mu$ plane that then shows the presence of a tricritical point joining
the lines of second order and first order chiral phase transitions, which we
are able to locate precisely.  In Sec. V we present other relevant
thermodynamical quantities and show explicitly the existence of a mixed chiral
symmetry restored/broken phase, the analogous of a liquid-gas phase in the
$P-1/\rho$ plane.  The entropy, latent heat, and other important quantities
are also evaluated.  In Sec. VI we present the next order results, at $T=0$
and $\mu=0$, that allows us to assess the convergence of the OPT in this
model.  Our conclusions are presented in Sec.  VII. Three appendices are
included to show some technical details and the renormalization for the
interpolated model up to second order.

\section{The Gross-Neveu Model}

The Gross-Neveu model is described by the Lagrangian density for a fermion
field $\psi_k$ ($k=1,\ldots,N$) given by \cite{gn}

\begin{equation}
{\cal L} =
\bar{\psi}_{k} \left( i \not\!\partial\right) \psi_{k} +
m_f {\bar \psi_k} \psi_k
+ \frac {g^2}{2} ({\bar \psi_k} \psi_k)^2\;,
\label{GN}
\end{equation}
where the summation over fermionic species is implicit in the above equation,
with e.g.  $\bar{\psi}_k \psi_k = \sum_{k=1}^N \bar{\psi}_k \psi_k$.  When
$m_f=0$, the theory is invariant under the discrete transformation
\footnote{Note that in $d=3$ this is only true if one considers $4\times 4$
  Dirac matrices.}

\begin{equation}
\psi \to \gamma_5 \psi \,\,\,,
\end{equation}
displaying a discrete chiral symmetry (CS).  {}For the studies of the model
Eq.  (\ref{GN}) in the large-$N$ limit it is convenient to redefine the
four-fermion interaction as $g^2 N = \lambda$. Since $g^2$ vanishes like $1/N$
we study the theory in the large-$N$ limit with fixed $\lambda$ (see, e.g.,
\cite{coleman}).

At finite temperature and density the model can be studied in terms of the
grand partition function given by
 
\begin{equation}
Z(\beta,\mu) = {\rm Tr} \exp\left[ - \beta \left( 
H - \mu Q \right) \right] \;,
\label{Zbetamu}
\end{equation} 
where $\beta$ is the inverse of the temperature, $\mu$ is the chemical
potential, $H$ is the Hamiltonian corresponding to Eq. (\ref{GN}) and $Q=\int
dx \bar{\psi}_k \gamma_0 \psi_k$ is the conserved charge. Transforming Eq.
(\ref{Zbetamu}) to the form of a path integral in the imaginary-time
(Euclidean) formalism of finite temperature field theory \cite{kapusta}, we
then have

\begin{equation}
Z(\beta,\mu) = \int \prod_{k=1}^N D \bar{\psi}_k D \psi_k
\exp\left\{ - S_{E}[\bar{\psi}_k,\psi_k] \right\} \;,
\label{func}
\end{equation}
where the Euclidean action reads

\begin{equation}
S_{E}[\bar{\psi}_k,\psi_k] =
\int_0^\beta d \tau \int dx \left[ \bar{\psi}_k
\left(\not\!\partial + \mu \gamma_0- m_f \right ) \psi_{k} -
\frac {\lambda}{2N} ({\bar \psi_k} \psi_k)^2 \right] \;,
\label{action}
\end{equation}
and the functional integration in Eq. (\ref{func}) is performed over the
fermion fields satisfying the anti-periodic boundary condition in Euclidean
time: $\psi_k(x,\tau) = - \psi_k (x,\tau + \beta)$.

\section{The effective potential for the interpolated theory}

Let us now turn our attention to the implementation of the OPT method within
the GN model.  Usually, when employing this approximation one starts by
performing a linear interpolation on the original model in terms of a
ficticious parameter $\delta$ (used only for bookkeeping purposes), which
allows for further expansions.  According to this OPT interpolation
prescription \cite{linear} (for a long, but far from complete list of
references on the method, see \cite{early}) the deformed four fermion theory
reads \cite{prdgn2}

\begin{equation}
{\cal L}_{\delta}(\psi, {\bar \psi}) =
\bar{\psi}_{k} \left( i \not\!\partial\right) \psi_{k} +
(1-\delta)\, \eta \, {\bar \psi_k} \psi_k
+ \delta \frac {\lambda}{2N} ({\bar \psi_k} \psi_k)^2\;.
\label{GNlde}
\end{equation}

\noindent
So, that at $\delta=0$ we have a theory of free fermions while at $\delta=1$
the original theory is reproduced.  Now, the introduction of an auxiliary
scalar field $\sigma$ can be achieved by adding the quadratic term,

\begin{equation}
- \frac{ \delta N}{2 \lambda} \left ( \sigma +
\frac {\lambda}{N} {\bar \psi_k} \psi_k \right )^2 \,,
\end{equation}
to ${\cal L}_{\delta}(\psi, {\bar \psi})$. We are then led to the interpolated
model

\begin{equation}
{\cal L}_{\delta} =
\bar{\psi}_{k} \left( i \not\!\partial\right) \psi_{k} -
\delta \sigma {\bar \psi_k} \psi_k - (1-\delta) \, \eta \, 
{\bar \psi_k} \psi_k
- \frac {\delta N }{2 \lambda } \sigma^2 + {\cal L}_{ct,\delta}  \;,
\label{GNdelta}
\end{equation}

\noindent
where ${\cal L}_{ct,\delta}$ is the part of Lagrangian density containing the
necessary counterterms for renormalization, whose coefficients are allowed to
be $\delta$ and $\eta$ dependent \cite{ldegn,prd1}.  As it is well known the
$2+1$-dimensional GN model is not renormalizable in the usual perturbative
expansion, but is renormalizable in the $1/N$ expansion \cite{Park}, which has
the property of modifying non-perturbatively the usual behavior under power
counting. Though our renormalization procedure in the OPT expansion is more
similar to a perturbative renormalization, this is not a real obstacle for our
analysis.  On general grounds it is always possible to calculate physical
quantities in a non-renormalizable model, at the price of introducing new
counterterms at successive orders, which simply means that the sensitivity to
an implicit cutoff of the model is expected to be more pronounced than in a
renormalizable theory \cite{Collins}. Such a procedure is commonly and
successfully applied in many effective theories, like e.g. typically in chiral
perturbation theory (for a recent review of chiral perturbation theory with
emphasize on the renormalization procedure see e.g.  \cite{chpt} and
references therein). However, concretely in our case, at first order of the
OPT expansion all the relevant quantities are actually finite (when using
dimensional regularization as we do here), thus completely unambiguous.  Next,
at second order, that we also investigate to some extent in this paper, it
turns out, that the only potentially non-renormalizable contributions to the
effective potential actually vanish, such that only standard (i.e. mass,
wave-function, etc) counterterms are necessary to cancel the divergences. It
should be mentioned however that this is somehow an accident of using
dimensional regularization, which therefore delays at most, {\it i.e.} to much
higher orders, the necessary introduction of new counterterms in our case.  A
detailed account for renormalization of the GN3d model in the OPT up to second
order, will be presented in Apps. B and C.

{}From the Lagrangian density in the interpolated form, Eq.  (\ref{GNdelta}),
we can immediately read the corresponding new {}Feynman rules in Minkowski
space. Each Yukawa vertex carries a factor $-i \delta$ while the (free)
$\sigma$ propagator is now $-i \lambda/(N \delta)$.  The LDE dressed fermion
propagator is

\begin{equation}
S_F(P)=\frac{i}{\not \! P - \eta_*+i\epsilon}\;,
\label{SF}
\end{equation}
where

\begin{equation}
\eta_*= \eta -(\eta - \sigma_c)\delta\;.
\label{eta*}
\end{equation}

Any quantity computed from the above rules, at some finite order in $\delta$,
is dependent on the parameter $\eta$, which then must be fixed somehow. Here,
as in most of the previous references on the OPT method, $\eta$ is fixed by
using the principle of minimal sensitivity (PMS). In the PMS procedure one
requires that a physical quantity $\Phi^{(k)}$, that is calculated
perturbatively to some $k$-th order in $\delta$, be evaluated at the point
where it is less sensitive to this parameter.  This criterion then translates
into the variational relation \cite{pms}

\begin{equation} 
\frac {d \Phi^{(k)}}{d \eta}\Big |_{\bar \eta, \delta=1} = 0 \;. 
\label{PMS} 
\end{equation} 

\noindent
The optimum value $\bar \eta$ that satisfies Eq. (\ref{PMS}) must be a
function of the original parameters, including the couplings, thus generating
non-perturbative results.

\subsection{The OPT Effective Potential}
\label{subseclargeNlde}

\begin{figure}[htb]
  \vspace{0.5cm}
  \epsfig{figure=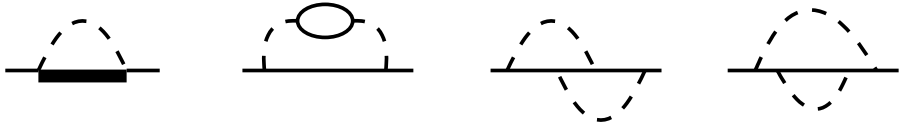,angle=0,width=12cm}
\caption[]{\label{graphs1} Diagrams contributing to the self energy to 
  order-$\delta^2$.  The thick continuous fermionic lines represent $\eta_*$
  dependent terms which must be further expanded, while the thin continuous
  lines represent $\eta$ dependent fermionic propagators and the dashed lines
  represent the $\sigma$ propagator.Diagrams (a) and (b) (of order-$\delta$
  and order-$\delta^2$, respectively) contribute with $1/N$, while diagrams
  (c) and (d) (both of order-$\delta^2$) contribute with $1/N^2$. Within
  dimensional regularization only graphs (b) and (c) are divergent.  Tadpole
  graphs are not shown as they do not contribute to the effective potential
  nor to the counterterms (in $d=2+1$) at the perturbative order we restrict
  ourselves to.}
\end{figure}

\begin{figure}[htb]
  \vspace{0.5cm}
  \epsfig{figure=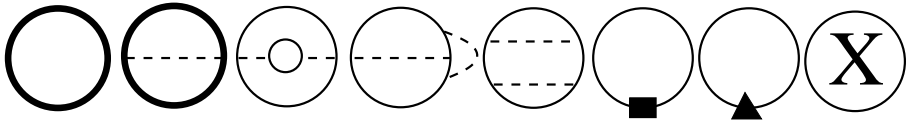,angle=0,width=12cm}
\caption[]{\label{graphs2} Diagrams contributing to $V_{\rm eff}/N$ to 
  order-$\delta^2$.  The thick continuous fermionic lines represent $\eta_*$
  dependent terms which must be further expanded while the thin continuous
  lines represent $\eta$ dependent fermionic propagators and the dashed lines
  represent the $\sigma$ propagator. The first (order-$\delta^0$) contributes
  with $1/N^0$, the second and third (order-$\delta$ and order-$\delta^2$
  respectively) contribute with $1/N$. The fourth and fifth (both of
  order-$\delta^2$) contribute with $1/N^2$. The sixth and seventh represent
  contributions due to the mass and wave function renormalization counterterms
  respectively. The last graph represents the zero point energy subtraction
  term. }
\end{figure}

The different contributions to the order-$\delta^2$ self-energy are displayed
in {}Fig. \ref{graphs1}.  We can use these self-energy terms to evaluate the
vacuum graphs contributing to the effective potential as shown in {}Fig. \ref
{graphs2}

In the sequel we make use of the following notations. The four-momentum $P$ is
given by $P = (P_0,{\bf p})$, where $P_0 = i(\omega_n - i\mu)$, with
$\omega_n= (2 n+1)\pi T$, $n=0,\pm 1,\pm 2,\ldots$, are the Matsubara
frequencies for fermions. The momentum integrals, when passing from
Minkowski to Euclidean space-time, we here denote by

\[
-i \int_p^{(T)} \equiv T \sum_{n=-\infty}^{+\infty} \int \frac{d^{d-1} p}{(2
  \pi)^{d-1}} \;,
\]
where all space momentum integrals are performed using dimensional
regularization, $d=3-\epsilon$. The renormalization procedure (which is only
necessary at order-$\delta^2$ since all relevant quantities are explicitly
finite at order-$\delta$) is carried out in the modified minimal subtraction
scheme ($\overline{\rm MS}$).

The order-$\delta$ OPT effective potential is obtained from the first two
diagrams shown in {}Fig. \ref{graphs2} and, using the previous Feynman rules
for the GN in the OPT, it is given by

\begin{equation}
\frac{V_{{\rm eff},\delta^1}(\sigma_c,\eta)}{N} = \delta
\frac {\sigma_c^2}{2 \lambda} +
i  \int_p^{(T)}  {\rm tr}\ln \left(\not \! P - \eta \right)+
\delta  i  \int_p^{(T)}   {\rm tr}
\frac {\eta-\sigma_c}{\not \! P - \eta + i \epsilon}
+ \frac {\Delta V_{\rm eff}^{(a)}}{N} \;,
\label{general}
\end{equation}
where $\Delta V_{\rm eff}^{(a)}/N$ brings the first $1/N$ correction to the
effective potential shown by the second diagram in {}Fig.  \ref{graphs2}. This
is given by \cite{root}

\begin{equation}
\frac{\Delta V_{\rm eff}^{(a)}}{N} =
- \frac {i}{2N}  \int_p^{(T)}
{\rm tr} \left [\frac {\Sigma_a(\eta)}{\not \! P - \eta +
i \epsilon} \right ]\;,
\label{VN1}
\end{equation}
where the trace is over Dirac's matrices only \footnote{The factor $-1$
  corresponding to a closed fermionic has already been taken into account
  \cite {root}. } while the term $\Sigma_a$ represents the first contribution
shown in {}Fig. \ref{graphs1} to the fermion self-energy,

\begin{equation}
\Sigma_a (\eta) = -\delta  \frac {\lambda}{N} 
i  \int_q^{(T)}  \frac {1}{\not \! Q - \eta+i \epsilon}\;.
\label{Sigma1}
\end{equation}
After taking the traces in Eq. (\ref{general}) and rearranging the terms one
obtains

\begin{eqnarray}
\frac{V_{{\rm eff},\delta^1}(\sigma_c,\eta)}{N} &= &\delta
\frac {\sigma_c^2}{2 \lambda} +
 2  i  \int_p^{(T)} \ln \left( P^2 - \eta^2 \right)+\delta
4 i  \int_p^{(T)}
\frac {\eta(\eta-\sigma_c)}{P^2 - \eta^2 + i \epsilon} \nonumber \\
&+& \delta  \frac{2 \lambda}{N} \eta^2\left [i  \int_p^{(T)}
\frac {1}{P^2 - \eta^2 + i \epsilon} \right ]^2 + \delta \frac{2 \lambda}{N}
\left [i  \int_p^{(T)} \frac {P_0}{P^2 - \eta^2 + i \epsilon}
\right ]^2 \,\,.
\label{vlde1ddim}
\end{eqnarray}
Then, at finite temperature and chemical potential, one finds (see appendix A
for the relevant integrals and Matsubara sums leading to this result)

\begin{eqnarray}
\frac{V_{{\rm eff},\delta^1}(\sigma_c,\eta)}{N} &= &\delta
\frac {\sigma_c^2}{2 \lambda} + \frac{ |\eta|^3}{3\pi} + 
\frac{|\eta|T^2}{\pi} I_1(a,b) + \frac{T^3}{\pi} I_2(a,b) \nonumber \\
&-&\delta \frac{\eta(\eta-\sigma)}{\pi}\left [ |\eta| + T I_3(a,b)\right ]
+ \delta \frac{\lambda \eta^2}{2(2\pi)^2N}
\left [ |\eta| + T I_3(a,b)\right ]^2 +
\delta \frac{\lambda T^4}{2(2\pi)^2N} [I_4(a,b)]^2 \;,
\label{vlde1d3}
\end{eqnarray}
where we have defined the functions

\begin{equation}
I_1(a,b)= {\rm Li}_2[-e^{-(a-b)}] + {\rm Li}_2[-e^{-(a+b)}] \,\,\,,
\label{I1}
\end{equation}

\begin{equation}
I_2(a,b)= {\rm Li}_3[-e^{-(a-b)}] + {\rm Li}_3[-e^{-(a+b)}] \,\,\,,
\label{I2}
\end{equation}

\begin{equation}
I_3(a,b)=\ln \left [ 1+ e^{-(a-b)} \right ] + 
\ln \left [ 1+ e^{-(a+b)}\right ] \,\,\,,
\label{I3}
\end{equation}

\begin{equation}
I_4(a,b)= {\rm sgn}(\mu)\left [ a \ln \left ( \frac{1+e^{a+b}}{1+e^{a-b}} 
\right )
+{\rm Li}_2[-e^{a+b}] - {\rm Li}_2[-e^{a-b}] \right ] \,\,\,,
\label{I4}
\end{equation}
with $a=|\eta|/T$ and $b=|\mu|/T$.

The $T\to 0$ limit for each of the elements appearing in Eq. (\ref{vlde1d3})
are:

\begin{eqnarray}
&& \lim_{T \to 0} T^2 I_1(a,b) = -\frac{1}{2}  
\left(|\mu| - |\eta|  \right)^2 \theta(|\mu| - |\eta|)\;, 
\label{I1T0}\\
&& \lim_{T \to 0} T^3 I_2(a,b) = \frac{1}{6} 
\left(|\eta| - |\mu|\right)^3 \theta(|\mu| - |\eta|) \;, 
\label{I2T0}\\
&&\lim_{T \to 0} T I_3(a,b) = \left( |\mu| - |\eta| \right) 
\theta( |\mu| - |\eta|)\;, 
\label{I3T0}\\
&&\lim_{T \to 0} T^2 I_4(a,b) = \frac{1}{2} {\rm sgn}(\mu)
\left(\eta^2 - \mu^2 \right) \theta( |\mu| - |\eta|) \;,
\label{I4T0}
\end{eqnarray}
where $\theta( |\mu| - |\eta|)$ is the step function.

\section{Optimization and Numerical Results Beyond Large-$N$}

Before proceeding to the specific $d=3$ case, considered in this work, let us
apply the PMS to the most general order-$\delta$ effective potential, which is
given by Eq. (\ref {vlde1ddim}).  This exercise will help the reader to
visualize the way the OPT-PMS resums the perturbative series. Setting
$\delta=1$ and applying the PMS to Eq. (\ref{vlde1ddim}) we obtain that

\begin{eqnarray}
&&\left \{ \left [ \eta - \sigma_c + \eta \frac {\lambda}{N}
\left (i  \int_p^{(T)}  \frac {1}{P^2 - \eta^2 + i \epsilon}
\right ) \right ] \left( 1 + \eta \frac{d}{d\eta} \right )
\left [i  \int_p^{(T)} \frac {1}{P^2 - \eta^2 + i \epsilon}
\right . \right ]\nonumber \\
& &+\left . \frac{\lambda}{N} \left (i  \int_p^{(T)}
\frac {P_0}{P^2 - \eta^2 + i \epsilon}  \right )\frac{d}{d \eta}
\left (i   \int_p^{(T)}  \frac {P_0}{P^2 - \eta^2 + i \epsilon}
\right ) \right \}\Bigr|_{\eta = {\bar \eta}} =0\,\,.
\label{pmsselfconsistent}
\end{eqnarray}
As one can see in App. A, Eq. (\ref{intP0p2}), the last term of the above
equation only survives when $\mu \ne 0$. In the case $\mu=0$, Eq.
(\ref{pmsselfconsistent}) factorizes in a nice way which allows us to
understand the way the OPT-PMS procedure resums the series producing
non-perturbative results.  With this aim one can easily check that (at
$\delta=1$)

\begin{equation}
\Sigma_a(\eta,T,\mu) = - \frac{\lambda}{N} \eta
\left [i  \int_p^{(T)} \frac {1}{P^2 - \eta^2 + i \epsilon}
\right ]  \;.
\end{equation}

\noindent
Then, when $\mu=0$, the PMS equation factorizes to

\begin{equation}
\left[ {\bar \eta} - \sigma_c - \Sigma_a({\bar \eta},T,\mu=0) \right]
\left( 1 + {\bar \eta} \frac{d}{d{\bar \eta}} \right )
\left [i  \int_p^{(T)} \frac {1}{P^2 - {\bar \eta}^2 + 
i \epsilon}
\right ]=0\;,
\label{selfconsistent}
\end{equation}
leading to the self-consistent relation

\begin{equation}
{\bar \eta} =\sigma_c + \Sigma_a({\bar \eta},T,\mu=0) \;,
\label{selfcons}
\end{equation}
which is valid for any temperature and number of space-time dimensions. In
this way the OPT fermionic loops get contributions containing $\sigma_c$ as
well as rainbow (exchange) type of self-energy terms, like the first graph of
figure \ref{graphs1}.  Note that when $N \to \infty$, ${\bar \eta}=\sigma_c$
and the large $N$ result is exactly reproduced \cite{prdgn2}. The mathematical
possibility

\begin{equation}
i  \int_p^{(T)} \frac {1}{P^2 -
{\bar \eta}^2 + i \epsilon} =0 \;,
\end{equation}
corresponds to the unphysical, coupling independent, solution discussed in
Ref. \cite{prdgn2}.  Note that in the $d=3$ case one obtains

\begin{equation}
\Sigma_a(\eta,T,\mu) = \frac{\lambda \eta}{4\pi N} 
\left[ |\eta| + TI_3(a,b) \right] \;.
\end{equation}

{}For numerical purposes, taking $\lambda \to -\lambda$ and defining $\Lambda=
\pi/|\lambda|$, one can consider the dimensionless PMS equation:

\begin{equation}
\left\{\left[ \eta - \sigma_c + \frac{\eta}{4N} \left(|\eta| +
T I_3(a,b) \right)\right] \left ( 1+ \eta \frac{d}{d \eta} \right )
\left( |\eta| +
T I_3(a,b) \right)+ \frac{T^4}{4N} I_4(a,b) \frac{d}{d \eta} I_4(a,b) 
\right \} \Bigr|_{\eta ={\bar \eta}} =0 \;.
\label{pmsdless}
\end{equation}
where $\eta,\sigma_c,T$, and $\mu$ are in units of $\Lambda$.

\subsection{The $T=0$ and $\mu=0$ case}

Let us start by analyzing each of the different possible cases involving the
temperature and chemical potential corrections.  {}For $T=0$ and $\mu=0$ we
have that, to order-$\delta$,

\begin{equation}
\frac{V_{\rm eff}^{\delta^1}(\eta,\sigma_c)}{N}=  
\delta \frac{\sigma_c^2}{2\lambda} + \frac {|\eta|^3}{3\pi} - 
\delta \frac{ \eta(\eta-\sigma_c) |\eta|}{\pi}  + \delta 
\frac {\lambda \eta^2 |\eta|^2}{2 (2\pi)^2 N} \;.
\label{vefftmuzero}
\end{equation}
Note that $V_{\rm eff}^{\delta^1}(\eta,\sigma_c)= V_{\rm
  eff}^{\delta^1}(-\eta,-\sigma_c)$ and by the virtue of this symmetry we
shall look for $\bar \eta(\sigma_c)$ only for $\sigma_c > 0$, since for
$\sigma_c < 0$ it is obvious that $\bar \eta(\sigma_c)= - \bar
\eta(-\sigma_c)$.

Then, using $\lambda \to -\lambda$ and $\Lambda= \pi/|\lambda|$, one can write
the free energy, at $T=0$ and $\mu=0$, as

\begin{equation}
\frac{V_{\rm eff}^{\delta^1}(\eta,\sigma_c)}{N }=
-\delta \frac{\sigma_c^2 \Lambda}{2\pi} + \frac {\eta^3}{3\pi}
- \delta \frac{ \eta^2(\eta-\sigma_c)}{\pi}
- \delta \frac { \eta^4 }{8 \pi N \Lambda}\;,
\label {vefftmuzero2}
\end{equation}
where the notation is consistent with the fact that we are only interested in
$\sigma_c > 0$ (in this case only $\eta > 0$ can recover the large-$N$ result
as $N \to \infty$ as can be seen from the PMS solution, Eq. (\ref
{selfcons})). Then $d V_{\rm eff}/d \sigma_c =0$ at $\sigma_c = {\bar
  \sigma}_c$ gives

\begin{equation}
{\bar \sigma}_c = \eta^2/\Lambda\;.
\label{sig1}
\end{equation}
At the same time the PMS equation $d V_{\rm eff}/d \eta =0$ at $\eta = {\bar
  \eta}$ gives the relation:

\begin{equation}
{\bar \eta}= \bar{\sigma}_c - \frac{{\bar \eta}^2}{4 N \Lambda}\;,
\end{equation}
from where we then obtain the expression

\begin{equation}
{\bar \eta}= {\bar \sigma_c} \,{\cal F}(N) \;,
\label{sigetarel}
\end{equation}
with the function ${\cal F}(N)$ defined as

\begin{equation}
{\cal F}(N) = 1 - \frac {1}{4N} \;.
\label{FN}
\end{equation}

\begin{figure}[htb]
  \vspace{0.5cm}
  \epsfig{figure=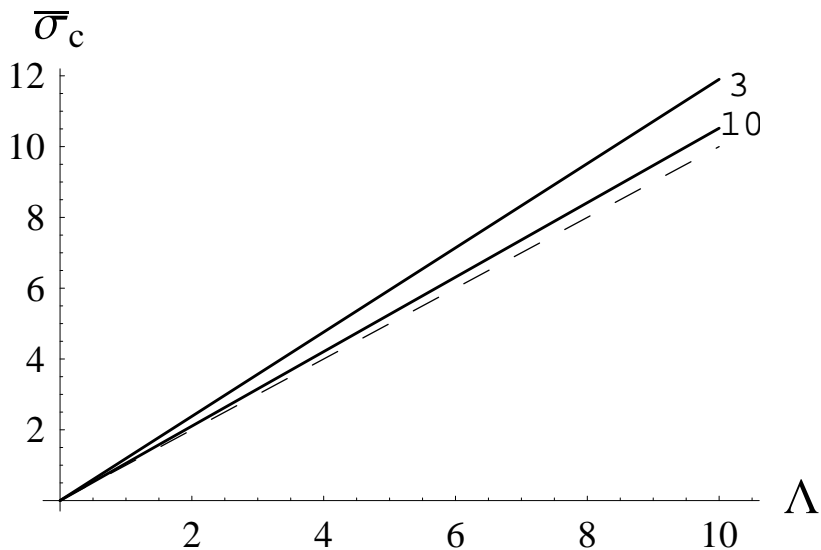,angle=0,width=12cm}
\caption[]{\label{gn3fig1} The dimensionless minimum ${\bar \sigma_c}$
  (in units of $\Lambda$) as a function of $\Lambda$ for $T=\mu=0$. The dashed
  line represents the $N \to \infty$ result, while the continuous lines were
  produced by the OPT-PMS at order-$\delta$. The numbers beside the curves
  identify the value of $N$ for each case.}
\end{figure}

\noindent
The above results then lead to the {\it optimized} value for the (dynamically
generated) vacuum expectation value for the scalar field , also shown in
\cite{letter},

\begin{equation}
{\bar \sigma_c} = \frac {\Lambda} {{\cal F}(N)^{2}} \,\,.
\label{sigrel}
\end{equation}
This result is contrasted with the large-$N$ result in {}Fig. \ref{gn3fig1}.

It is instructive to recall a similar result for the $d=2$ where the large-$N$
result is ${\bar \sigma_c}^N(\lambda)= M \exp(-\pi/\lambda)$ \cite{Park},
while the OPT result is ${\bar \sigma_c}^{\delta^1}= {\bar
  \sigma_c}^N(\lambda^*)/[1-1/(2N)]$, where $\lambda^*= \lambda[1-1/(2N)]$
\cite{prdgn2}.  The same happens here except that we have a factor 4 inside
the function dependent on $N$, Eq. (\ref{FN}), instead of a factor 2 found in
the $d=2$ case. This is because of the $4\times 4$ Dirac matrices considered
in the three dimensional problem here.  Thus, we have ${\bar
  \sigma_c}^{\delta^1}= {\bar \sigma_c}^N(\lambda^*)/[1-1/(4N)]$, recalling
that ${\bar \sigma_c}^N(\lambda) = \Lambda = \pi/|\lambda|$. {}Finally, when
we evaluate the thermodynamical potential in the sequel, it will be useful to
consider the optimized free energy at its minimum, $\sigma_c = {\bar
  \sigma}_c$, which is given by

\begin{equation}
\frac{V_{\rm eff}({\bar \eta},{\bar \sigma}_c)}{N} = - 
\frac {\Lambda^3}{6 \pi{\cal F}(N)^3}\;.
\label{veffmin}
\end{equation}

\subsection{The $T \ne 0$ and $ \mu=0$ case}

Next, let us consider the case $T \ne 0$ and $\mu=0$ when the free energy can
be written, using again $\lambda \to -\lambda$, and $\Lambda = \pi /|\lambda|$
as:

\begin{eqnarray}
\frac{V_{\rm eff}^{\delta^1}(\eta,\sigma_c, T)}{N }&=&  
-\delta \Lambda \frac{\sigma_c^2}{2\pi} + \frac {\eta^3}{3\pi} + 
\frac{2}{\pi} \left [ \eta T^2{\rm Li}_2 \left(-e^{-\eta/T} \right) + 
T^3{\rm Li}_3 \left( -e^{-\eta/T}\right) \right ] \nonumber \\
&-& \frac{\eta}{\pi}(\eta-\sigma_c)
\left [ \eta + 2T \ln \left( 1+e^{-\eta/T} \right) \right ]
- \delta \frac{\eta^2}{8\pi N \Lambda}
\left [ \eta + 2T \ln \left( 1+e^{-\eta/T} \right) \right ]^2 \,\,,
\label{veffT}
\end{eqnarray}
where we have considered again the case $\sigma_c > 0$ and $\eta >0$. From the
result given in Eq. (\ref{selfcons}), ${\bar \eta} = \sigma_c +
\Sigma_a(\bar{\eta},T,\mu=0)$, we immediately obtain the self-consistent
temperature dependent relation

\begin{equation}
{\bar \eta}= \sigma_c - \frac{\bar \eta}{4N \Lambda}\left [ {\bar \eta} +
2T \ln(1+e^{-{\bar \eta}/T})\right ] \,\,.
\label{pmstmuzero}
\end{equation}
It is a simple matter to apply $d V_{\rm eff}/d \sigma_c =0$ at $\sigma_c =
{\bar \sigma}_c$ to Eq.  (\ref{veffT}) to obtain

\begin{equation}
{\bar \sigma}_c = \frac{\eta}{\Lambda}\left [ \eta + 2T \ln(1+e^{-\eta/T})
\right ]\;,
\label{sigbartmuzero}
\end{equation}
which can be used in Eq. (\ref{pmstmuzero}) to yield ${\bar \eta} = {\bar
  \sigma} {\cal F}(N)$ that, when inserted into Eq. (\ref {sigbartmuzero}),
allows to study the thermal behavior of order parameter (${\bar \sigma}_c(T)$)
via the extremum of $V_{\rm eff}^{\delta^1}(\eta,\sigma_c, T)$, given by
\cite{letter}

\begin{equation}
{\bar \sigma}_c(T){\cal F}(N) = \frac {\Lambda}{{\cal F}(N)} -
2T \ln [1+ e^{- {\bar \sigma}_c(T) {\cal F}(N)/T}]\;.
\end{equation}
{}From the above equation one retrieves the result ${\bar \sigma}_c(0)=
\Lambda {\cal F}(N)^{-2}$, as obtained in the previous subsection.  The
critical temperature $T_c$ for chiral symmetry restoration is obtained by
requiring that ${\bar \sigma}_c(T=T_c) =0$, which gives the result

\begin{equation}
T_c^{\delta^1} = \frac {\Lambda}{2\ln 2} \frac{1}{{\cal F}(N)}\;.
\label{Tc1}
\end{equation}

\begin{figure}[htb]
  \vspace{0.5cm}
  \epsfig{figure=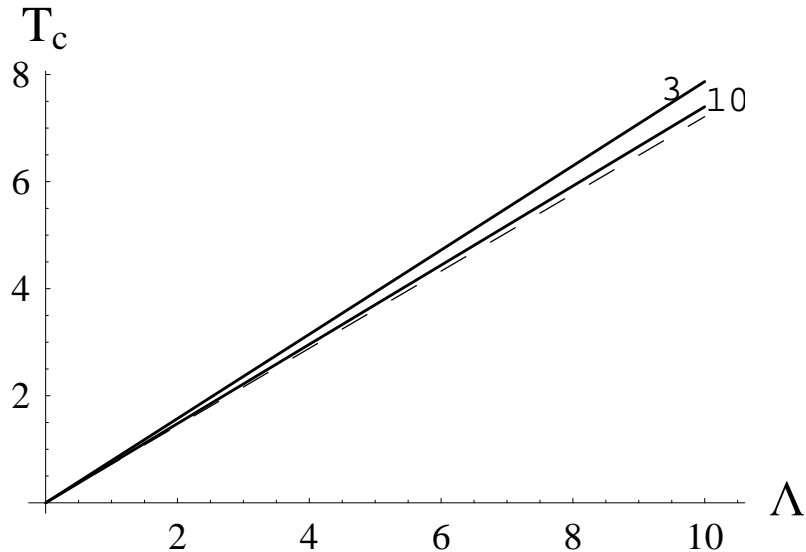,angle=0,width=12cm}
\caption[]{\label{tclde1} The  OPT critical temperature (at $\mu=0$ and
  in units of $\Lambda$) as a function of $\Lambda$. The dashed line
  represents the $N \to \infty$ result, while the continuous lines were
  produced by the OPT-PMS procedure at order-$\delta$.  The numbers beside the
  curves identify the value of $N$ for each case.}
\end{figure}

\noindent
This analytical result is shown in the {}Fig. \ref {tclde1}.  Note that a
numerical application of the PMS to the OPT effective potential, using Eqs.
(\ref{pmsdless}) and (\ref {veffT}), exactly reproduces the analytical result
Eq. (\ref{Tc1}).  Recall the similarity with the $d=2$ case \cite{prdgn2},
where the LDE result was obtained from the large $N$ result by the
replacement: $\lambda \to \lambda^*= \lambda [1 - 1/(2N)]$, just like the same
as in obtaining the scalar field vev, as discussed in the previous subsection.

\begin{figure}[htb]
  \vspace{0.5cm}
  \epsfig{figure=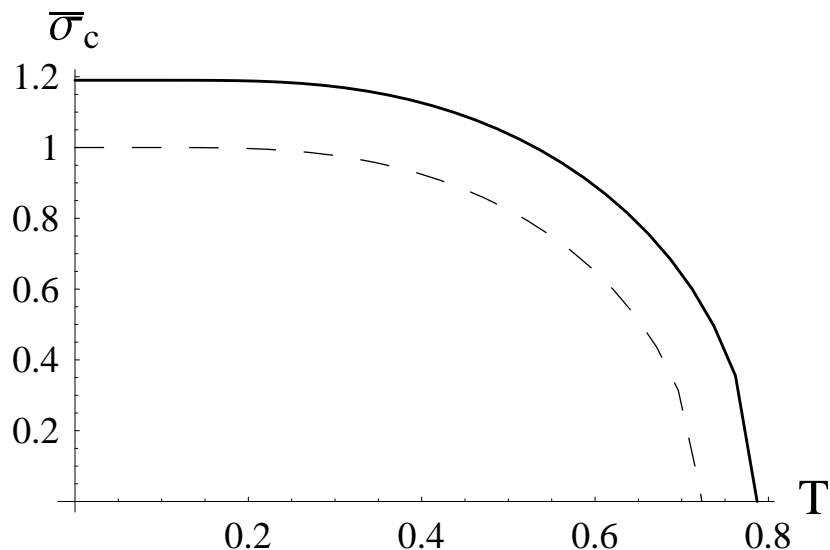,angle=0,width=12cm}
\caption[]{\label{ldephase} The large-$N$ (dashed line) and the OPT
  (continuous line) predictions for ${\bar \sigma}_c(T)$ at $N=3$.  All
  quantities are in units of $\Lambda$. The figure displays a continuous,
  second order, transition line. }
\end{figure}

Note also that, contrary to the $d=2$ case, our prediction for $T_c$ is always
greater (for finite $N$) than the large-$N$ prediction.  The transition is
found to be of the second kind, as illustrated by {}Fig. \ref{ldephase}.

\subsection{The $T= 0$ and $ \mu \ne 0$ case}

Let us now consider the case $T= 0$ and $ \mu \ne 0$.  {}From the general
expression of the effective potential at the first OPT order, Eq.
(\ref{vlde1ddim}), and using the $T\to 0$ results shown in Eqs.
(\ref{I1T0})--(\ref{I4T0}), we find that the chemical potential dependent
effective potential is given by

\begin{eqnarray}
\frac{V_{\rm eff}^{\delta^1}(\eta,\sigma_c,\mu, T=0)}{N} &=&   
\delta \frac{\sigma_c^2}{2\lambda}
+ \frac{|\eta|^3}{3 \pi} - \frac{\delta \, (\eta- \sigma_c)\, 
\eta}{\pi} |\eta|+
\frac{\delta \lambda \eta^4}{2 (2 \pi)^2 N}   
\nonumber \\
&+&
 \left[ \frac{1}{2 \pi} \left( - \frac{2}{3} |\eta|^3 + |\mu| \eta^2 - 
\frac{|\mu|^3}{3}
\right) -  \frac{\delta \eta(\eta- \sigma_c)}{\pi} 
\left( |\mu| - |\eta| \right) +
\frac{\delta \lambda \eta^2}{2 (2 \pi)^2 N} \left(\mu^2 - \eta^2 \right)
\right.
\nonumber \\
&+& \left. \frac{\delta \lambda }{8 (2 \pi)^2 N} 
\left(\eta^2 - \mu^2 \right)^2 \right] \theta(|\mu| - |\eta|) \;.
\label{vlde1mu}
\end{eqnarray}
{}For $\mu=0$, obviously, Eq. (\ref{vlde1mu}) reduces to Eq.
(\ref{vefftmuzero}).

To evaluate the critical value $\mu_c$ for chiral symmetry restoration, it is
sufficient to compare the values of the effective potential at the minimum
$V_{\rm eff}(\bar\sigma_c, T=0)$ with its value for $\mu \ne 0$ for $\bar
\sigma_c =0$. This is from the same line of reasoning employed in the $d=2$
case discussed in Ref. \cite{prdgn2}. In this case, we obtain the point where
the two minima of the effective potential, at $\bar{\sigma}_c=0$ and at
$\bar{\sigma}_c \neq 0$ and $\mu=\mu_c$ coincide, {\it i.e.}, there is a value
$\mu_c$ which satisfies

\begin{equation}
V_{\rm eff}^{\delta^1}(\bar\sigma=0, \mu=\mu_c, T=0) = 
V_{\rm eff}^{\delta^1}(\bar\sigma_c,  \mu=0,T=0)\;.
\label{mucdef}
\end{equation}
It is a simple algebraic exercise to calculate both members of this equality.
We first obtain from Eq. (\ref{vlde1mu}) that

\begin{equation}
\frac{V_{\rm eff}^{\delta^1}(\sigma=0, \mu_c, T=0)}{N} =
-\frac{1}{6\pi} |\mu_c|^3 \left(1 -3 \frac{\lambda}{16\pi N} |\mu_c| \right)\;.
\label{Vmutzero}
\end{equation} 

\noindent
Then, to evaluate the right hand side of Eq. (\ref{mucdef}) we use Eq.
(\ref{vefftmuzero}) together with the relation between $\bar\sigma_c$ and
$\bar\eta$ in Eq. (\ref{sigetarel}), {\it i.e.}, $\bar\sigma_c =
\bar\eta/[{\cal F}(N)]$, which gives

\begin{equation}
\frac{V_{\rm eff}^{\delta^1}(\bar\sigma_c, \mu =0, T=0)}{N}
= |\bar\eta|^3 \left( \frac{1}{2 {\cal F}(N)} -\frac{2}{3}
+\frac{\lambda \bar\eta}{8\pi N} \right)
=  -\frac{|\bar\eta|^3}{6\pi}\;,
\label{Vsigcmuzero}
\end{equation}

\noindent
where the last simplification arises from using the definition of ${\cal
  F}(N)$, Eq. (\ref{FN}), and noting that $\bar\eta = -\pi/[\lambda\,{\cal
  F}(N)]$.  Note thus that $V_{\rm eff}^{\delta^1}(\bar\sigma_c, \mu = T=0)$
has formally the same simple expression as the leading order one, except of
course that it includes non-trivial $1/N$ corrections via the explicit
expression of $\bar\eta$. Now we may compare Eqs. (\ref{Vmutzero}) and
(\ref{Vsigcmuzero}) to finally extract $\mu_c$ \cite{letter},

\begin{equation}
|\mu_c| =\frac{ \Lambda }{{\cal F}(N)}\;\left(1+ \frac{3}{16N}\:
\frac{|\mu_c|}{\Lambda} \right)^{-1/3} \;,
\label{muctzero}
\end{equation}
where we used again $\Lambda = \pi /|\lambda|$.  {}For $N=3$, we find from Eq.
(\ref{muctzero}) the solution,

\begin{equation}
\frac{\mu_c}{\Lambda}\simeq 1.06767 \;,
\end{equation}
which agrees with the numerical results obtained in the next subsection.

\subsection{The $T\ne 0$ and $ \mu \ne 0$ case}

\begin{figure}[htb]
  \vspace{0.5cm}
  \epsfig{figure=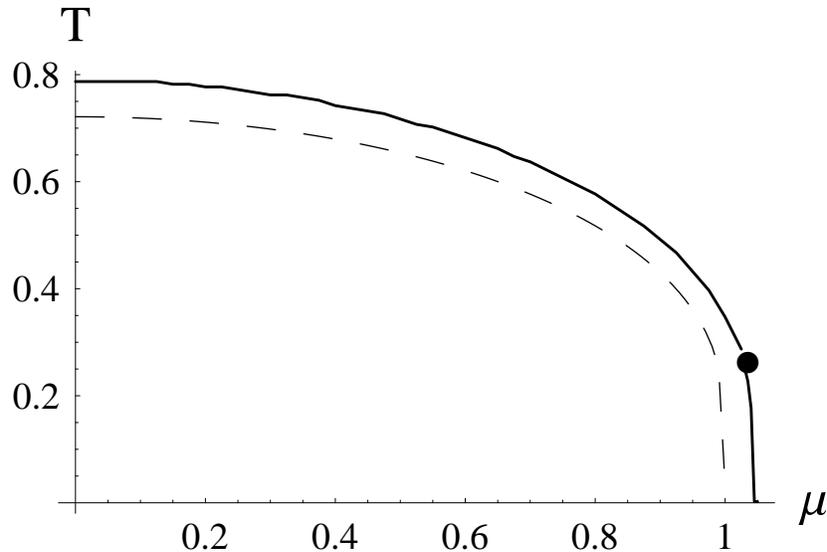,angle=0,width=12cm}
\caption[]{\label{gn3fig2} The large-$N$ (dashed line) and OPT
  (continuous line) predictions for the phase diagram at $N=3$. All quantities
  are in units of $\Lambda$. The black dot indicates the position of a
  tricritical point, located at $T_{\rm tcr} \simeq 0.251$ and $\mu_{\rm tcr}
  \simeq 1.029$.  Below this point the transition is of the first kind while
  above the point it is of the second kind.}
\end{figure}

{}Finally, turning now for the case of both finite temperature and finite
chemical potential, we obtain the full phase diagram of the three dimensional
GN model.  The numerical application of the PMS shows how the phase diagram is
qualitatively and quantitatively affected by finite $N$ corrections.  {}Figure
\ref{gn3fig2} shows the situation for $N=3$. The
CSB region is augmented with respect to the large-$N$ predictions, when
expressed 
in units of our reference scale $\Lambda \equiv \pi/|\lambda|$. This is
clear also from our results for $\bar{\sigma}_c$, $T_c$ and $\mu_c$.  This
appears 
at first sight in contrast with the lattice results \cite{hands}, which show a
decreasing of the CSB region at finite $N$, as compared to the large-$N$
results.  However, whithin our approximation, the increase of the size of 
the CSB region is rather small,
being about 5$\%$ for $N=3$, while the increase would be of only about 2$\%$
for the $N=12$ case considered in Ref. \cite{hands}, which in turn predicts a decrease of
about 10$\%$.  One should moreover 
note that in Ref. \cite{hands} the authors present their results for the phase
diagram with quantitites normalized by the scalar vacuum expectation value
obtained from the lattice simulations. In our case it means that from Eqs.
(\ref{sigrel}), (\ref{Tc1}) and (\ref{muctzero}), taking $N=3$ for
instance, $T_c/\bar{\sigma}_c \simeq 0.661$ and $\mu_c/\bar{\sigma}_c \simeq
0.897$, while for $N=12$, $T_c/\bar{\sigma}_c \simeq 0.707$ and
$\mu_c/\bar{\sigma}_c \simeq 0.976$, which approximately agrees with the results presented in Ref. 
\cite{hands} within their level of precision. Note also that our results
agree reasonably with recent analysis of the $2+1$ GN model from exact renormalization
group methods \cite{exrg} (at least within the errors quoted there,
and for the case of vanishing wave function of the auxiliary scalar field, which
is the appropriate comparison to our analysis).
Thus the reduction or increase with respect to large $N$ results appears to be just a matter
of scaling. In the present work we consider the scale defined as $\Lambda$ (the
vacuum expectation value of the scalar field at large-$N$) as more appropriate
to present the results. In a previous work
\cite{prdgn2}, we have considered the same model but in 1+1 dimensions using
exactly the same approximation. There, our result for the phase diagram showed
a drastic change concerning the size of the CSB which was about 30 $\%$ {\it
  smaller} than the one produced by the large-$N$ approximation for $N=3$.
Now, in 2+1 dimensions, it turns out that the CSB region predicted by the OPT
is very close to the one predicted by the large-$N$ approximation.  In
summary, in the fixed normalization scale used ($\Lambda$), it looks like the
OPT predicts a drastic decrease in the size of the CSB region in 1+1
dimensions whereas in 2+1 dimensions it seems to support, at least at lowest
OPT order, the CSB size (as well as the numerical values of $\bar \sigma_c$,
$T_c$ and $\mu_c$) predicted at large-$N$. We shall see in next section that
a partial investigation of higher OPT order corrections (for $T=\mu=0$)
indicates a good stability of these first order results. 
On the other hand, the nature of
the transition line predicted at large-$N$ in 1+1 dimensions is unaffected by
the OPT at order-$\delta$ while it drastically changes in 2+1 dimensions,
as we now start to discuss.

Concerning the nature of the transition lines shown in {}Fig. \ref{gn3fig2},
recall that the large-$N$ approximation predicts that it is of the second kind
everywhere, except at $T=0$ where it suddenly becomes of the first kind at a
critical value of $\mu= \mu_c^N = \Lambda$. Using lattice Monte Carlo
simulations for the GN model, Kogut and Strouthos \cite{kogutmc} concluded
that, for $N=4$, there should be a tricritical point on the section of the
phase boundary defined by $T/T_c^N \le 0.230$ and $\mu/\mu_c^N \ge 0.970$. Our
evaluations predict that the observed first order transition at $T=0$ spreads
out through the transition line until it reaches a tricritical point at
$T_{\rm tcr} \simeq 0.251 \, \Lambda$ and $\mu_{\rm tcr} \simeq 1.029 \,
\Lambda$ (for $N=3$).

\begin{figure}[htb]
  \vspace{0.5cm}
  \epsfig{figure=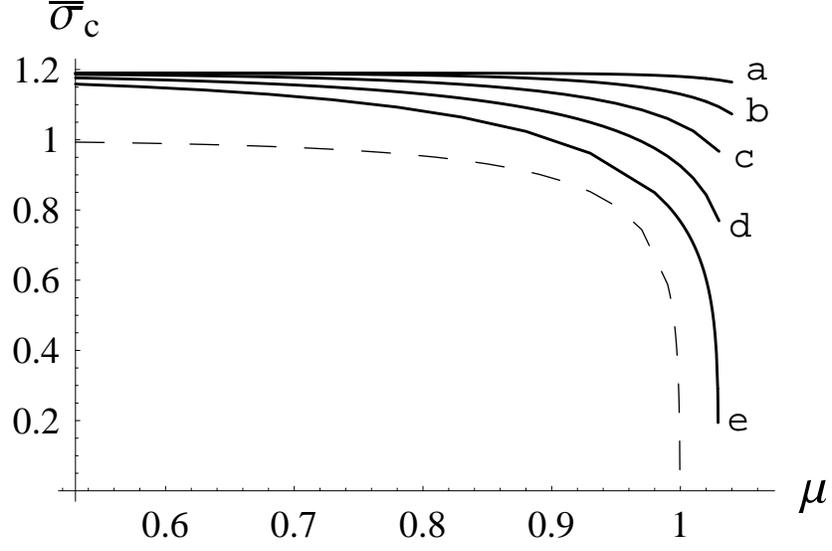,angle=0,width=12cm}
\caption[]{\label{envelope}  The order parameter, ${\bar \sigma}_c$,
  as a function of $\mu$ for different temperatures. The continuous lines
  represent the OPT results for $N=3$ and the labels on the figure represent
  the different temperatures: $T_a = 0.050, T_b=0.100, T_c=0.150, T_d=0.200$
  and $T_e=0.250$. All these values are smaller than the tricritical value
  $T_{\rm tcr}=0.251$ and the associated curves clearly display first order
  transitions. All quantities are given in units of $\Lambda$.  {}For
  reference, the figure also shows the large-$N$ result (dashed line) for
  $T=0.150$.}
\end{figure}

{}Figure \ref{envelope} shows an envelope of curves that displays how the
abrupt first order transition at $T=0$ becomes smoother as the temperature
increases. One observes that the discontinuity gap becomes smaller as the
temperature approaches the tricritical value, $T_{\rm tcr}= 0.251 \, \Lambda$.

\begin{figure}[htb]
  \vspace{0.5cm}
  \epsfig{figure=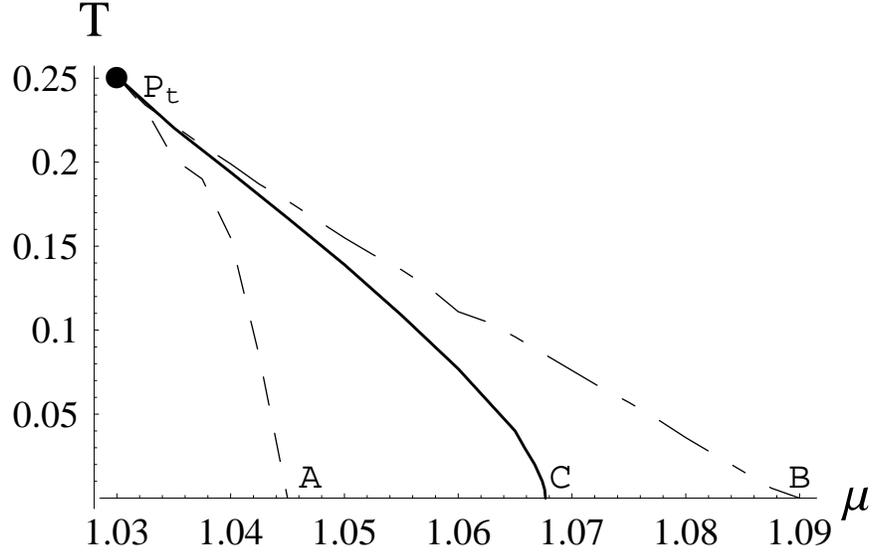,angle=0,width=12cm}
\caption[]{\label{metalines}  Part of the phase diagram that corresponds 
  to the metastable region (reproduced here from \cite{letter}).  The dashed
  line joining points $P_t$ and $A$ refers to the development of a minimum at
  the origin. The continuous line linking the tricritical point to $C$ ($T=0$
  and $\mu=\mu_c$) is the first order transition line, while the dot-dashed
  line joining $P_t$ to $B$ is the second metastability line and signals that
  the minima that occur away from the origin have disappeared.}
\end{figure}

It is also important to analyze the occurrence of metastability lines related
to the first order transition line ($T < T_{\rm tcr}$) in {}Fig.
\ref{gn3fig2}. With this aim we offer {}Fig. \ref{metalines}, where the dashed
line joining the tricritical point, $P_t$, to point $A$ corresponds to the
appearance of a minimum at $\sigma_c=0$, whereas the continuous line joining
points $P_t$ and $\mu_c$ refers to the first order transition line. {}Finally,
the dot-dashed line joining points $P_t$ and $B$ refers to the vanishing of
the minima that occurs away from the origin. It is important to note how small
the metastable region $A-P_t-B$ is. As a matter of fact, point $A$ occurs at a
value of $\mu$ which is about $2 \%$ smaller than $\mu_c$, whereas $B$ occurs
at a value which is about $3 \% $ greater than $\mu_c$. In $d=2$ these values
are of about $30 \%$ (see Ref. \cite {italianos}).

\begin{figure}[htb]
  \vspace{0.5cm}
  \epsfig{figure=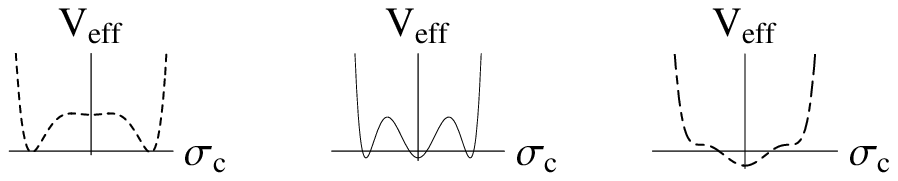,angle=0,width=12cm}
\caption[]{\label{metapots} The shape of the free energy corresponding 
  to the metastable region. The left panel shows the situation corresponding
  to the dashed $P_t-A$ line shown in {}Fig. \ref{metalines}, while the middle
  and right panels correspond to lines $P_t-C$ and $P_t-B$ respectively. }
\end{figure}

The reader may visualize the three situations shown in {}Fig. \ref{metalines}
by examining the form of the free energy shown in {}Fig. \ref{metapots}.

\section{The liquid-gas phase}

We are now in position to perform a more physical interpretation of our
results by examining other relevant thermodynamical quantities. The
thermodynamical potential, $\Omega(\mu,T)$, for instance, is related to the
free energy at its minimum, $\sigma_c = {\bar \sigma}_c$. It is given by

\begin{equation}
\Omega(\mu, T) = V_{\rm eff} ({\bar \eta}, {\bar \sigma}_c, \mu, T)\;.
\end{equation}
Note that we have defined this quantity in terms of the {\it optimized} free
energy. This is an important remark because we are considering different
physical quantities and one could wonder which one to optimize. Here, our
choice is to optimize the free energy since all other thermodynamical
quantities may be obtained from it. It is usual to normalize the
thermodynamical potential by subtracting a ``bag'' term, ${\cal B}$, given by
${\cal B} = \Omega (0,0)$, so that the pressure as well as the energy density
vanish at $T=0$ and $\mu=0$. In view of Eq. (\ref {veffmin}) the bag term is
simply given by

\begin{equation}
{\cal B} =- \frac {\Lambda^3}{6 \pi{\cal F}(N)^3} \,\,.
\label{bag}
\end{equation}
Then, the {\it normalized} thermodynamical potential is just $\Omega_N(\mu,T)
= \Omega(\mu,T) - {\cal B}$. At the same time, the (normalized) pressure is
given by $P(\mu,T) = - \Omega_N(\mu,T)$ from which one may obtain the density

\begin{equation}
\rho = \frac {\partial P}{\partial \mu} \,\,,
\end{equation}
and the entropy density

\begin{equation}
{\cal S} =\frac {\partial P}{\partial T} \,\,.
\end{equation}
{}Finally, the (normalized) energy density is given by ${\cal E}= - P + T{\cal
  S} + \mu \rho$.  Recalling that, due to the gap equation, $\partial
P/\partial {\bar \sigma}_c =0$ and that, due to the PMS equation, $\partial
P/\partial {\bar \eta} =0$ one obtains the density

\begin{equation}
\rho = 
-\frac{{\bar \eta} T^2}{\pi} I_{1,\mu} - \frac{T^3}{\pi} I_{2,\mu} 
+ \frac{(\bar{\eta}-\bar{\sigma}_c)\, \bar{\eta}}{\pi} T I_{3,\mu}
- \frac{\lambda \bar{\eta}^2}{(2\pi)^2N} T
\left( \bar{\eta} + T I_3\right) I_{3,\mu}  -
\frac{\lambda T^4}{(2\pi)^2N} I_4 I_{4,\mu} \;,
\label{rho}
\end{equation}
where $I_{i,\mu} \equiv \partial I_i/\partial \mu$, and $I_i$, $i=1,\ldots,4$,
are given by the function in Eqs. (\ref{I1}) - (\ref{I4}). At the same time
one obtains that the entropy density is given by

\begin{eqnarray}
{\cal S} &=& 
- 2 \frac{\bar{\eta} T}{\pi} I_1 -\frac{{\bar \eta} T^2}{\pi} I_{1,T}  
-3 \frac{T^2}{\pi} I_2 - \frac{T^3}{\pi} I_{2,T}  
+ \frac{(\bar{\eta}-\bar{\sigma}_c)\, \bar{\eta}}{\pi}  I_{3}
+  \frac{(\bar{\eta}-\bar{\sigma}_c)\, \bar{\eta}}{\pi} T I_{3,T}
\nonumber \\
&-& \frac{\lambda \bar{\eta}^2}{(2\pi)^2N}
\left( \bar{\eta} + T I_3\right) (I_{3} + T I_{3,T}) -
\frac{\lambda T^3}{(2\pi)^2N} (2 I_4^2 + T I_{4,T}) \;,
\label{Sd}
\end{eqnarray}
where $I_{i,T} \equiv \partial I_i/\partial T$.

\begin{figure}[htb]
  \vspace{0.5cm}
  \epsfig{figure=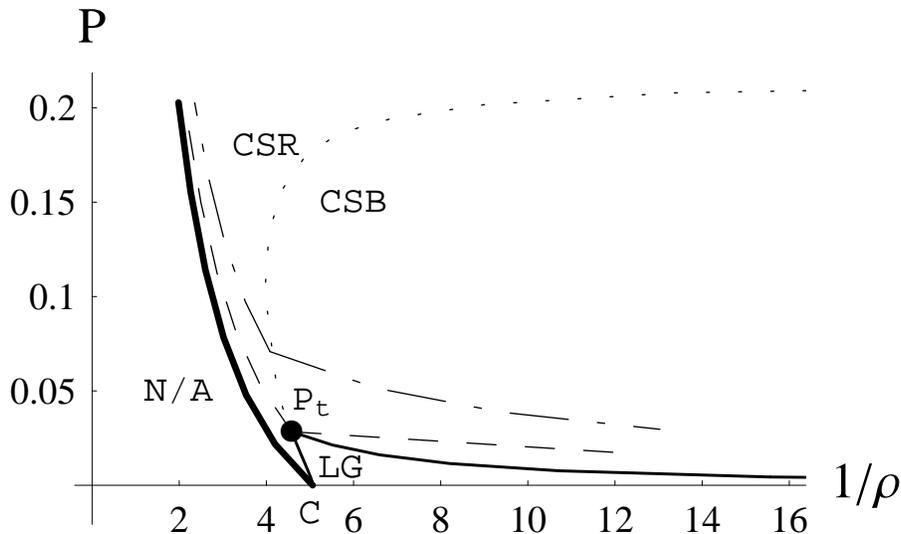,angle=0,width=12cm}
\caption[]{\label{figPrho} The phase diagram in the $P$-$1/\rho$ plane 
  for $N=3$. The thick continuous line is the $T=0$ isotherm which limits the
  region accessible to the system. The region labeled by $N/A$ is not
  accessible. The dotted line is the mapping of the second order transition
  line. The chiral symmetric region (CSR) corresponds to the ``liquid'' phase
  while the region where chiral symmetry is broken (CSB) corresponds to the
  ``gas'' phase. The $LG$ region starting at the the tricritical point, $P_t$,
  is limited by first order transition lines and corresponds to the mixed
  ``liquid-gas'' phase. Point $C$ ($P=0$ and $\rho \simeq 0.197 \, \Lambda^2$)
  corresponds to $T=0, \mu=\mu_c$.  The isotherm represented by the dashed
  line corresponds to the tricritical temperature, $T_{\rm tcr} = 0.251 \,
  \Lambda$ while the dot-dashed line represents the isotherm corresponding to
  $T=0.397 \, \Lambda$. The pressure, $P$, is in units of $\Lambda^3$ while
  the density, $\rho$, is given in units of $\Lambda^2$.  }
\end{figure}

Having all the above quantities, we can now analyze, for instance, the phase
diagram in the physically more accessible $P-1/\rho$ plane as shown by {}Fig.
\ref{figPrho}. This figure displays one of our most important results which
indicates that a mixed ``liquid-gas" phase, previously unknown to exist within
this model, develops at low pressure values.  Three isotherms are shown and
the one corresponding to $T=0$ defines the edge of the region accessible to
the system. The dotted line above the tricritical point is just the mapping of
the corresponding second order transition line in the $T-\mu$ plane. Note that
in the $P-1/\rho$ plane the first order transition line displayed in the
$T-\mu$ plane (corresponding to $T<T_{\rm tcr}$) splits in two parts
corresponding to the value of the density at the two degenerate minima which
produce identical pressure values. Not being able to determine the existence
of the mixed phase, Kogut and Strouthos argued that the liquid-gas transition
was either (i) extremely weak, (ii) very close to the chiral transition, or
(iii) not realized in this model \cite {kogutmc}. Our result shown in {}Fig.
\ref {figPrho} suggests that their second hypothesis was the correct one.
Moreover, these authors had used $N=4$, that was the smallest number allowed
by the hybrid Monte-Carlo algorithm used in their simulations. With this
number the tricritical point appears at an even lower value of $P$ (actually,
for $N \to \infty$ it happens at $P=0$) and was consequently harder to be
detected within their approximation.

\begin{figure}[htb]
  \vspace{0.5cm}
  \epsfig{figure=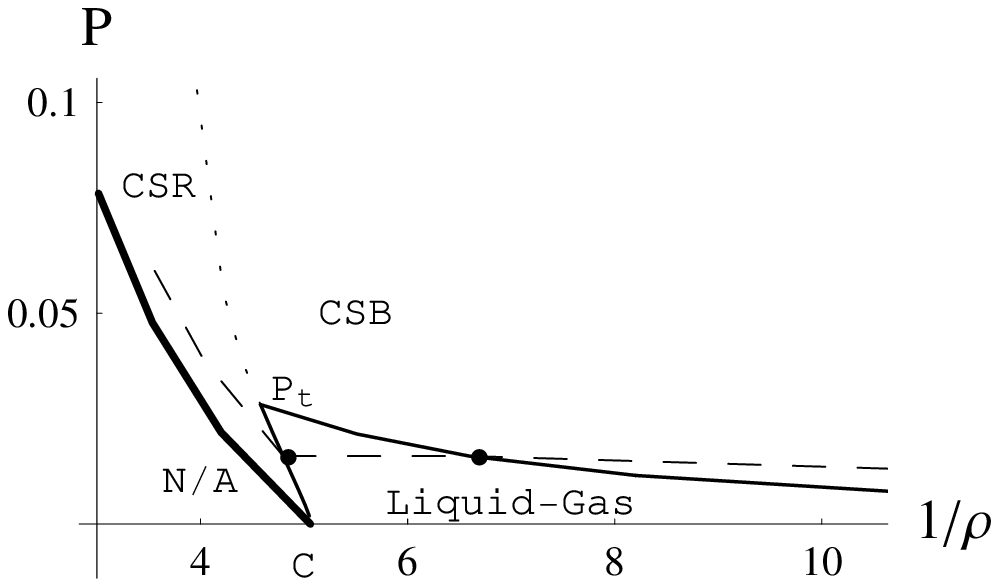,angle=0,width=12cm}
\caption[]{\label{figPrhozoom} Detail of the liquid-gas phase in the 
  $P$-$1/\rho$ plane for $N=3$.  The isotherm represented by the dashed line
  corresponds to a temperature $T=0.194 \, \Lambda$ that is smaller than the
  tricritical temperature, $T_{\rm tcr} = 0.251 \, \Lambda$ . The two dots are
  joined by a straight line in a Maxwell construction.  The pressure, $P$, is
  in units of $\Lambda^3$ while the density, $\rho$, is given in units of
  $\Lambda^2$.  }
\end{figure}

{}Figure \ref{figPrhozoom} shows a detailed view of the ``liquid-gas" phase
seen in {}Fig. \ref{figPrho} displaying how an isotherm whose temperature is
smaller than the tricritical value crosses the mixed phase region. The
horizontal line in the ``coexistence" region was drawn by connecting the value
of the pressure at the boundaries, corresponding to ``mixed" states. This
picture observes the Maxwell construction which derives from the equality of
the chemical potentials at the edge of the two phases.

\begin{figure}[htb]
  \vspace{0.5cm}
  \epsfig{figure=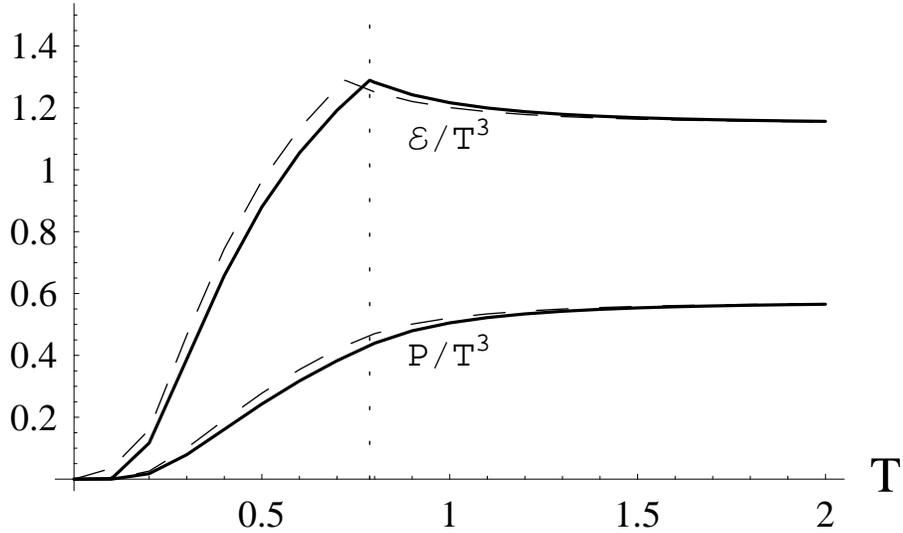,angle=0,width=12cm}
\caption[]{\label{ptcubomuzero} The dimensionless quantities ${\cal E}/T^3$ 
  and $P/T^3$ as functions of the temperature for $\mu = 0$. The continuous
  lines are the OPT results for $N=3$, while the dashed lines represent the
  large-$N$ results. The vertical dotted line is the OPT critical temperature
  for $N=3$.  }
\end{figure}

{}Figure \ref{ptcubomuzero} shows ${\cal E}/T^3$ and $P/T^3$ as functions of
the temperature for $\mu=0$. Note that for high temperatures ${\cal E}/T^3 \to
-3 \zeta(3)/\pi \simeq 1.14$, while $P/T^3 \to -3 \zeta(3)/(2\pi) \simeq 0.57$
(where $\zeta(3) \simeq 1.202$), as one can guess by looking at the equations
for ${\cal E}$ and $P$ at $\mu=0$.  In those high temperature regimes ($T >
T_c$), we have ${\bar \sigma}_c =0$ and ${\bar \eta} \to 0$. At high
temperatures both curves are symmetrical with respect to the numerical value $
\sim 0.85$.

\begin{figure}[htb]
  \vspace{0.5cm}
  \epsfig{figure=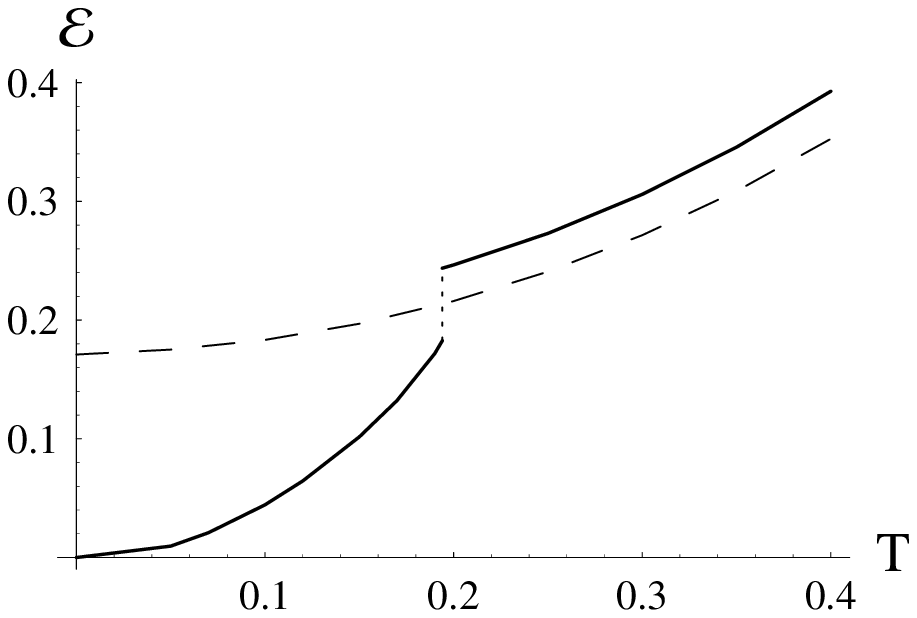,angle=0,width=12cm}
\caption[]{\label{enedenvst} The energy density, ${\cal E}$, as a 
  function of the temperature for $\mu=1.040 > \mu_{\rm tcr}=1.029$ (for
  $N=3$). Both quantities are in units of $\Lambda$. The continuous line is
  the OPT result and shows the presence of latent heat signaled by the
  discontinuity at $T_c (\mu=1.040 \Lambda)= 0.194 \Lambda$.  The large-$N$
  result is represented by the dashed line and the discontinuity happens at
  $T=0$. Both ${\cal E}$ and $T$ are in units of $\Lambda$.  }
\end{figure}

Let us now check for the presence of latent heat, which is inherent to first
order phase transitions. We do this by examining the energy density as a
function of the temperature. {}For this, one chooses a value of $\mu$ that
corresponds to the first order transition, like, for example, any $\mu$ such
that $\mu_c > \mu > \mu_{\rm tcr}$.  Recall that for the case $N=3$, $\mu_{\rm
  tcr}=1.029 \Lambda$ and $\mu_c = 1.067\, \Lambda$, so we choose, without
loss of generality, the value $\mu=1.040 \Lambda$. One then expects to see a
discontinuity in the line corresponding to ${\cal E}(T)$ at $T=T_c(\mu=0.140
\Lambda)= 0.194 \Lambda$. This is indeed the case as shown in {}Fig.
\ref{enedenvst}. The same figure shows the large-$N$ result, where the
discontinuity happens only at $T=0$, which can be understood by recalling that
within this approximation the first order phase transition happens only at the
point $T=0$ and $\mu = 1.000 \Lambda$.

\section{ Order-$\delta^2$ Results at $T=0$ and $\mu=0$}

Let us now investigate the order-$\delta^2$ contributions that are given by
the three-loop graphs shown in {}Fig. \ref {graphs2}.  Actually, a complete
evaluation of these graphs at finite $T$ and $\mu$ turns out to be very
cumbersome, so we shall restrict ourselves in the present work to the $T=0$
and $\mu=0$ case which is more tractable.  This will at least allow us to have
a reasonable quantitative estimate of the expected higher order corrections to
our previous order-$\delta$ results. We plan to tackle the calculation of the
full $T$ and $\mu$ dependence in a future work.

\subsection{The order $\delta^2$ three-loop contribution to the free energy}

The total order $\delta^2$ three-loop contribution can be easily extracted
from Ref. \cite {root} and reads

\begin{equation}
\frac{\Delta V_{\rm eff}^{(b,c,d)}}{N} =
- \frac {i}{4N}  \int \frac{d^d p}{(2\pi)^d} \:
{\rm tr} \left [\frac {\Sigma_b(p,\eta) + \Sigma_c(p,\eta)+\Sigma_d(p,\eta)}
{\not \! p - \eta +
i \epsilon} \right ]\;,
\label{VN2}
\end{equation}
where $\Sigma_i(p,\eta)$, $i=b,c,d$, correspond to panels (b),(c), and (d)
(second, third and forth diagrams, respectively) in {}Fig. \ref{graphs1}.

The most complicated contributions arise from the first and second terms of
the above equation (or equivalently corresponding to the third and fourth
graphs of Fig. \ref{graphs2}) since in this case the self energies depend on
the momentum $p$.  After taking the traces, etc and using dimensional
regularization with $d=3-\epsilon$, the corresponding integral to be evaluated
reads

\begin{equation}
\frac{V_{\rm eff}^{(b,c)}}{N}=  \delta^2 \lambda^2 \frac{4}{N} 
\left ( 1- \frac{1}{4N} \right )
\left(\frac{e^{\gamma_E}M^2}{4\pi} \right)^{3\epsilon/2} i
\int \frac{d^d p}{(2 \pi)^d}
\left\{ i \int \frac{d^{d} k}{(2 \pi)^{d}}
\frac{k^2 + pk + \eta^2}{(k^2-\eta^2)[(p+k)^2 -\eta^2]}
\right\}^2\;.
\end{equation} 
Introducing next appropriate {}Feynman parameters to disentangle the different
momenta integrations, we obtain the following expression

\begin{eqnarray}
\frac{ \Delta V_{\rm eff}^{(b,c)}}{N}&=&  - \delta^2 \lambda^2
\frac{\eta^5}{(4\pi)^3
(4\pi)^{3/2}} \frac{4}{N} \left ( 1- \frac{1}{4N} \right ) 
\Gamma\left(-\frac{5}{2}+\frac{3\epsilon}{2}\right)(2-\epsilon)^2 
\left( \frac {e^{\gamma_E}
M^2}{\eta^2} \right )^{3\epsilon/2} \nonumber \\
&\times& \int_0^1 d \alpha \, d\beta \, d\gamma \,
g(\gamma) H(\alpha,\beta,\gamma) \,\,\,,
\label{3loopdiv}
\end{eqnarray}
where $M$ is the arbitrary $\overline{\rm MS}$ renormalization scale,

\begin{equation}
g(\gamma)= [\gamma(1-\gamma)]^{-3/2 + \epsilon/2} \,\,\,,
\end{equation}
and

\begin{equation}
H(\alpha,\beta,\gamma) = \left [ \gamma\: \alpha(1-\alpha) + 
(1-\gamma)\: \beta (1-\beta) \right ]^{-3/2 +\epsilon/2} \,\,.
\label{defH}
\end{equation}
The evaluation of the final integrals over {}Feynman parameters in Eq.
(\ref{3loopdiv}) is rather technical and details of this calculation are left
to appendix B.  One arrives at the final result for the third and fourth
three-loop diagrams of {}Fig. \ref{graphs2} as given by

\begin{equation}
\frac{\Delta V_{\rm eff}^{(b,c)}}{N} =  \delta^2 \lambda^2
\frac{\eta^5}{30 \pi^3}
\frac{4}{N} \left ( 1- \frac{1}{4N} \right )
\left[\frac{1}{\epsilon} +\frac{41}{10} 
- 4 \ln 2 - \frac{X}{2} - 3 \ln \frac{\eta}{M} \right]\;.
\label{fingraph1}
\end{equation}
where $X\sim 1.63669$ is a numerical constant obtained from the integrations.
Next, the self-energy entering the last term of Eq. (\ref{VN2}), corresponding
to the last diagram shown in {}Fig. \ref{graphs1}, that leads to the
contribution to $V_{\rm eff}$ shown by the fifth diagram in {}Fig.
\ref{graphs2}, it is a tadpole like graph that can be easily evaluated with
the standard {}Feynman rules. It is finite and gives the contribution,

\begin{equation}
\frac{\Delta V_{\rm eff}^{(d)}}{N}=-
\delta^2 \lambda^2 \frac{\eta^5}{2 N^2 (4\pi)^3} \;.
\label{fingraph2}
\end{equation}

At the three-loop order, the free energy contains divergent terms that lead to
the $1/\epsilon$ term shown in Eq. (\ref{fingraph1}).  The renormalization is
performed as usual, by introducing the appropriate counterterms (mass, wave
function, etc).  Actually the perturbative two-loop fermion self-energy in
Fig. 14 exhibits divergent terms of non-renormalizable kind, with higher power
of momentum dependence, as expected since the model is not perturbatively
renormalizable. Although it would not be a problem in principle to treat those
new divergences with appropriate counterterms, similarly to what is done in
other effective theories, it turns out that these non-renormalizable
counterterms do not contribute to the three-loop free energy in dimensional
regularization, so that only standard mass, wave-function and vacuum
counterterms are needed in practice to render the effective potential at
${\cal O}(\delta^2)$ finite.  Thus the renormalization as performed in the
$\overline{\rm MS}$ scheme after dimensional regularization does not introduce
new parameters at the three-loop order, for the quantities we are interested
in.  The arbitrariness of the final physical result for the three-loop
effective potential simply takes the standard form of a (logarithmic)
dependence on an arbitrary renormalization scale.  The details of the
calculation of these counterterms contributions are discussed in App. C. Now,
by adding all contributions, including the finite ones arising from the
counterterms (see App. C), the renormalized three-loop effective potential can
be cast into the form:

\begin{equation}
\frac{\Delta V_{\rm eff,ren}^{(b,c,d)}}{N} =  \delta^2 \lambda^2 \frac{\eta^5}
{\pi^3}
\left[ \frac{4}{N} \left ( 1- \frac{1}{4N} \right )
\left(\frac{3}{25} - \frac{X +7\ln 2}{60} 
- \frac{1}{12} \ln \frac{\eta}{M}\right) -\frac{1}{128\,N^2}
\right]\;.
\label{Eren3}
\end{equation}

\subsection{Optimization results at $T=0$ and $\mu=0$ at order $\delta^2$}

{}From the result for the ${\cal O}(\lambda^2\delta^2)$ three-loop
contribution, Eq. (\ref{Eren3}), we can now perform the $\delta$ expansion and
PMS optimization at this next order, limited however here to the special case
$T=0$ and $\mu=0$.  {}Following the same line of reasoning as in Sec. IV.A
above, care is to be taken by noticing that other $\delta^2$ terms are
generated by the appropriate expansion of $\eta^*$, as defined in Eq.
(\ref{eta*}), within the first order ${\cal O}(\delta \lambda)$ terms.
Explicitly, after having redefined $\lambda \to -\pi/\Lambda$ as previously,
we arrive at the expression,

\begin{eqnarray}
\frac{V_{\rm eff}^{\delta^2}(\eta,\sigma_c)}{N } &=&
-\delta \frac{\sigma_c^2 \Lambda}{2\pi} + \frac {\eta^3}{3\pi}
- \delta \frac{ \eta^2(\eta-\sigma_c)}{\pi}
- \delta \frac { \eta^4 }{8 \pi N \Lambda}
+ \delta^2 \frac{ \eta (\eta-\sigma_c)^2}{\pi}
+ \delta^2 \frac{ \eta^3 (\eta-\sigma_c)}{2\pi\,N\,\Lambda}
\nonumber \\
&+&
\delta^2  \frac{\eta^5}{\Lambda^2 \pi}
\left[ \frac{4}{N} \left ( 1- \frac{1}{4N} \right )
\left(\frac{3}{25} - \frac{X+7 \ln 2}{60}
- \frac{1}{12} \ln \frac{\eta}{M}\right) -\frac{1}{128\,N^2}
\right] \;.
\label{v2tmuzero}
\end{eqnarray}
The remaining is simply an algebraic exercise to apply the PMS procedure to
Eq. (\ref{v2tmuzero}). Similarly to the first order case, the gap equation, $d
V_{\rm eff}/d \sigma_c =0$ at $\sigma_c = {\bar \sigma}_c$ defines
$\bar{\sigma}_c$ as function of $\bar{\eta}$, while the PMS equation $d V_{\rm
  eff}/d \eta =0$ at $\eta = {\bar \eta}$ gives a further relation between
$\bar{\sigma}_c$ and $\bar \eta$.  However, at this next order, both relations
are more complicate, in particular the PMS equation is non-linear and involves
a $\ln(\eta)$ term. It is most convenient to use the gap equation, which gives

\begin{equation}
\bar\sigma_c = \frac{\eta^2}{2\eta-\Lambda}\:
\left( 1+\frac{\eta}{2N\,\Lambda} \right)\;,
\label{sigma2}
\end{equation}

\noindent
generalizing Eq. (\ref{sig1}), at second $\delta$-order, into the PMS
equation, which defines an equation depending only on $\bar\eta(N)$ (note that
the dependence upon $\Lambda$ can be simply factored out, e.g., by rescaling
$\bar\eta$ in units of $\Lambda$).  {}For instance, for the particular case
$N=3$, which is just sufficient for our illustration, one obtains the PMS
equation as

\begin{equation}
 \hat \eta^2 \left\{ 
 -0.0217944 (\hat \eta -0.819029) (\hat\eta +1.28009) 
 \left[ 3.48261 + \hat\eta (\hat\eta -3.6924)\right]
 - 0.162102 \left(\hat\eta-\frac{1}{2}\right) \hat\eta^2\: 
 \ln \frac{\hat\eta\Lambda}{M} \right\} = 0\;,
\label{etapms2}
\end{equation}
where we have defined, for convenience, the dimensionless mass parameter $\hat
\eta \equiv \eta/\Lambda$.  Eq. (\ref{etapms2}) can be solved numerically, to
find a non-trivial value of $\bar\eta$.  {}For this we have to set the
arbitrary renormalization scale $M$ in (\ref{etapms2}), originating from the
logarithmic dependence in Eq.  (\ref{v2tmuzero}), to some appropriate value. A
physically natural choice is to set $M=\Lambda$, that corresponds to the basic
scale and scalar vacuum expectation value  in the large-$N$ limit.  
An interesting feature of the
optimization result in the present case, is that the presence of the
$\ln\,(\eta)$ dependence in Eq. (\ref{etapms2}) largely reduces the number of
optimized solutions (usually a drawback of the PMS).  Indeed, one sees in Eq.
(\ref{etapms2}) that, without the logarithmic term, four different non-zero
(real or complex) $\bar \eta$ solutions would occur \footnote{One may wonder
  if the peculiar choice of the arbitrary scale: $M = \bar\eta$, thus
  canceling the logarithmic term, would not re-introduce the PMS
  multi-solution problem. But this gives no consistent optimal solutions since
  all other $\bar\eta$ solutions contradict this value of $\bar\eta$. This
  incidentally shows that Eq. (\ref{etapms2}) does not always have real
  solutions for any values of $M$.}.  In contrast, we find here numerically a
unique (real) solution (for $M=\Lambda$ and $N=3$):

\begin{equation}
\bar \eta \simeq 0.867\:\Lambda \;.
\label{etabar2}
\end{equation}
Next, we can just plug in this result into Eq.  (\ref{sigma2}), to obtain

\begin{equation}
\bar \sigma_c^{\delta^2} \simeq 1.1719\:\Lambda\;,
\label{sigmabar2}
\end{equation}
which appears to be very close to the first order result, ${\bar
  \sigma_c}^{\delta^1} \simeq 1.1901 \Lambda$, obtained from Eq.
(\ref{sigrel}) for $N=3$.  One may however question if this is an artifact of
our choice of arbitrary renormalization scale $\Lambda=M$.  Though this value
of the scale appears to be very natural, it is easy to study the impact of
varying it in a reasonable range around this value, like is sensible to do in
similar renormalization scale dependence studies in other theories (and which
give a rough estimate of higher order corrections). We can, e.g., vary $M$ in
the range $0.50 \Lambda \lesssim M \lesssim 1.50\Lambda$, which
correspondingly changes ${\bar \sigma}_c$ from values $1.16 \Lambda \lesssim
\bar{\sigma}_c \lesssim 1.38 \Lambda $ (note that for $M$ values too far apart
from $M\sim \Lambda$, Eq. (\ref{etapms2}) has no real solutions). {}From this
we conclude that the higher $\delta^2$ order corrections for $T,\mu =0$ to the
previous analysis are quite small, though their scale dependence is apparent
and can become non negligible. Actually, it is clear that the scale dependence
is essentially determined by the relative size of the $\ln (\eta/M)$ term with
respect to constant terms in the optimization Eq.  (\ref{etapms2}), which
turns out to be moderate \footnote{ It is interesting to note also that from
  Eq. (\ref{sigma2}) $\bar{\sigma}_c(\eta)$ has a minimum, ${\bar \sigma}_c
  \sim 1.16\,\Lambda$, at $\eta \sim 0.94 \,\Lambda$, independent of the scale
  $M$, thus ${\bar \sigma}_c$ cannot reach values below $\sim 1.16\Lambda$:
  when varying the scale $M$ to values lower than $\Lambda$, the optimal
  solution $\bar\eta$ of Eq. (\ref{etapms2}) decreases, so that ${\bar
    \sigma}_c(\bar\eta)$ will start to increase again.  Also, the unwelcome
  singularity of ${\bar \sigma}_c$ for $\eta=\Lambda/2$, apparent in Eq.
  (\ref{sigma2}), is not a solution of Eq. (\ref{etapms2}), so this singular
  value is never reached by the consistent PMS optimal solution $\bar\eta$.}.
On general grounds, renormalization scale dependence is expected to be rather
pronounced at lowest loop orders and damped by higher order perturbative
contributions \cite{Collins}. But in our framework the three-loop expression,
Eq. (\ref{Eren3}), is the very first perturbative order at which
renormalization, and thus scale dependence, occur for the effective potential,
so that the behavior is more similar to a one-loop quantity with respect to
renormalization scale dependence.  {}For the natural scale choice $M=\Lambda$,
however, these second order corrections for ${\bar \sigma}_c$ value in Eq.
(\ref{sigmabar2}) are very small, about only $1.5 \%$ of the first order
value, so in other words the OPT expansion seems to converge rather quickly.

\section{Conclusions}

In the present paper we have applied the alternative analytical
non-perturbative optimized perturbation theory (OPT) approach, through which
one can easily include finite $N$ effects, to a four-fermion theory with
discrete chiral symmetry represented by the massless Gross-Neveu model in 2+1
dimensions. We have then evaluated the free energy, or effective potential,
for the model at both finite temperature and finite chemical potential.  The
evaluation of the optimal value for this quantity has allowed us to derive and
study in detail other thermodynamical quantities, such as the pressure,
density, entropy and energy density.

Our main results in this paper include the analytical derivation of
expressions, going beyond the standard large-$N$ results, for the scalar field
vacuum expectation value, the critical temperature, and for critical chemical
potential related to chiral symmetry breaking/restoration. We have also
demonstrated the existence of a tricritical point, not seen in the large-$N$
approximation and the corresponding existence of a mixed chiral
restored-broken phase in the system for finite values of chemical potential
and temperature.  Concerning the phase diagram we recall that in a lattice
simulation, Kogut and Strouthos \cite{kogutmc} have predicted the existence of
tricritical point in the $T-\mu$ plane that is missed by the large-$N$
approximation.  However, within the numerical precision of their simulations,
those authors were unable to give its exact location. Here, we have not only
confirmed the existence of such a point but have also demonstrated how it can
be located for any value of $N$.  The formalism employed in this work has
allowed us to easily draw the first order transition line together with the
metastability lines.  This exercise showed that the metastable region is
rather small which possibly explains why its observation was not possible in
Refs. \cite {hands,kogutmc}.  Going to the $P-1/\rho$ plane has generated
another important result for the GN3d model.  Namely, the prediction of a
``liquid-gas" phase transition, so far unknown to exist in this model. All
these results drastically change the large-$N$ picture of the GN3d phase
diagram in which only a ``superconducting" phase and a ``normal" phase appear
\cite {rose}.  Although in our study the complete $T$ and $\mu$ dependence
could only be studied at the first non-trivial order of the $\delta$
expansion, we were able to estimate the higher order-$\delta^2$ corrections at
least at $T=0$ and $\mu=0$.  These corrections turn out to be reasonably small
(about only $1.5\%$ for a natural choice of renormalization scale). This also
indicates that all others OPT order-$\delta^2$ results are not expected to
distance too much from the computed order-$\delta$ results obtained for the
GN3d model.  So we may argue that all our phase transition and thermodynamical
results obtained in Secs. V and VI are likely to remain qualitatively
unaltered by higher order corrections in this framework.

We notice that some of our quantitative results (most notably those displayed
by {}Figs. 3 to 6) appear at first sight different from the overall 
behavior obtained with some
Monte Carlo simulations \cite{hands} that predict a reduced size for the CSB
region, while our results display an increase of the CSB region with respect
to the large N (equivalently mean field) results. As discussed
previously, as far as we can compare this seems to be only an artefact of 
the different reference scales chosen to express
the critical quantitites (e.g. the authors of Ref. \cite{hands} express 
the phase diagram in
terms of the scalar field vacuum expectation value, while we express all the
relevant quantities in units of the fixed reference scale ($\Lambda$).
But we note also that even for our choice of reference scale the increase with
respect the large-$N$ results is rather very small. This is opposite to what
is seen when the same approximation is applied to the 1+1 dimensional GN case
\cite{prdgn2} where a large decrease of the CSB region is observed compared to
the large-$N$ (mean-field) results. In this respect, the OPT applied to the 2+1
dimensions GN model leads to results closer to the mean-field approximation,
though the character of CSB/CSR transition line is qualitatively very
different.

As usual, within the OPT, all the large-$N$ results are exactly recovered from
our expressions by taking the limit $N \to \infty$.  In addition to the
explicit thermodynamical analysis presented here, we have also shown in
details, in Apps. B and C, the renormalization of the effective potential up
to ${\cal O} (\delta^2)$ in the OPT. A comparison to the case of
renormalization within the $1/N$ expansion is also briefly provided.  In the
case of the main expressions derived in this work, all results to ${\cal O}
(\delta)$ are shown to be finite within the dimensional regularization in the
OPT.  Divergences start to appear at second order.  At this order the
effective potential can be rendered completely finite just with standard
counterterms. Possible extensions of the present work are presented in Ref.
\cite {letter}.

\acknowledgments

M.B.P. and R.O.R. are partially supported by CNPq-Brazil. E.S. is partially
supported by CAPES-Brazil. R.O.R.  also acknowledges partial support from
FAPERJ. J-L.K. acknowledges partial support from UFSC.

\appendix

\section{Summing Matsubara frequencies and related formulas}

In this appendix we give the results for the main integrals and Matsubara sums
appearing along the text.  {}Following the general procedure for evaluating
these sums \cite{kapusta} and using dimensional regularization in the
$\overline{\rm MS}$ scheme, the momentum space integrals are written as

\[
\int \frac {d^{2} p}{(2 \pi)^{2}} \to \left(\frac{e^{\gamma_E} M^2}{4 \pi}
\right)^{\epsilon/2} \int \frac {d^{d-1} p}{(2 \pi)^{d-1}} \;,
\]
where $d=3-\epsilon$, $M$ is an arbitrary mass scale and $\gamma_E \simeq
0.5772$ is the Euler-Mascheroni constant.

{}For the general case of $T \ne 0$ and $\mu \ne 0$, we then obtain, for
example, that

\begin{eqnarray}
i\int_p^{(T)} {\rm ln} (P^2 - \eta^2 + i\epsilon) &=& \frac{|\eta|^3}{3 \pi}
+ \frac{\eta}{\pi} T^2 \left\{ {\rm Li}_2 \left[ - e^{-(|\eta| - |\mu|)/T}
\right] +  {\rm Li}_2 \left[ - e^{-(|\eta| + |\mu|)/T}
\right] \right\} 
\nonumber \\
&+& \frac{T^3}{\pi} \left\{ {\rm Li}_3 \left[ - e^{-(|\eta| - |\mu|)/T}
\right] +  {\rm Li}_3 \left[ - e^{-(|\eta| + |\mu|)/T}
\right] \right\}  \;,
\label{intln}
\end{eqnarray}

\begin{equation}
i\int_p^{(T)} \frac{1}{P^2 - \eta^2+ i\epsilon} = -\frac{T}{4\pi} 
\left\{  \frac{|\eta|}{T} + \ln \left [ 1+ e^{-(|\eta| - |\mu|)/T}
\right] +  \ln \left[ 1+ e^{-(|\eta| + |\mu|)/T}\right] \right\}
\;,
\label{intp2}
\end{equation}

\begin{eqnarray}
i\int_p^{(T)} \frac{P_0^2}{P^2- \eta^2+ i\epsilon }& =&
- i\,  {\rm sgn}(\mu) \frac{T^2}{4\pi} \left\{ \frac{|\eta|}{T} \ln \left[
\frac {1+ e^{(|\eta| + |\mu|)/T}}{1+ e^{(|\eta| - |\mu|)/T}} \right] +
{\rm Li}_2\left[-e^{(|\eta| + |\mu|)/T}\right] \right . \nonumber \\
&-& \left . {\rm Li}_2\left[-e^{(|\eta| - |\mu|)/T}\right]
\right\}\;,
\label{intP0p2}
\end{eqnarray}
where ${\rm sgn}(\mu)$ is the sign function and ${\rm Li}_\nu(z)$ is the
polylogarithm function and it is defined (for $\nu >0$) as \cite{Abramowitz}

\[
{\rm Li}_\nu(z)= \sum_{k=1}^{\infty} \frac{z^k}{k^\nu}\;.
\]

\section{Evaluation of three-loop free energy graphs at $T=0$ and $\mu=0$.}

In this appendix we give some technical details on the evaluation of the two
complicated contributions to the three-loop free energy. These are the third
and fourth graphs of Fig. 2 respectively, which give the expression Eq.
(\ref{3loopdiv}). There are probably different ways to perform this integral.
For convenience we choose to first integrate on the Feynman parameter
$\gamma$, which can be done analytically, with the result:

\begin{eqnarray}
\label{3loopab}
\frac{\Delta V_{\rm eff}^{(c,d)}}{N} &=& 
-\frac{4}{N}\left(1-\frac{1}{4N} \right)\frac{\delta^2\lambda^2\eta^5}
{(4\pi)^{9/2}} \:
\left(\frac{e^{\gamma_E} M^2}{\eta^2}\right)^{\frac{3\epsilon}{2}} \,
\Gamma\left(-\frac{5}{2}+\frac{3\epsilon}{2}\right) (2-\epsilon)^2
\nonumber \\  
& \times & 
\frac{\Gamma^2\left(-\frac{1}{2}+\frac{\epsilon}{2}\right)}
{\Gamma(-1+\epsilon)}\:
\int_0^1  d\alpha \, d \beta \,  
[\beta (1-\beta)]^{-\frac{3}{2}+\frac{\epsilon}{2}} \,
{_2 F_1} [-\frac{1}{2}+\frac{\epsilon}{2}, \frac{3}{2}-\frac{\epsilon}{2}, 
-1+\epsilon; 1-z] \;,
\end{eqnarray}
where $_2 F_1$ is the hypergeometric function, with $z \equiv \alpha
(1-\alpha)/[\beta(1-\beta)]$. Next we can use some well-known properties of
the hypergeometric function \cite{Abramowitz}, relating $_2 F_1[...;z]$ to $_2
F_1[...;1-z]$, and the power expansion in $z$ of the hypergeometric function
as

\begin{equation}
_2 F_1[p,q,r;z] = \sum_k \frac{(p)_k (q)_k}{(r)_k}\;\frac{z^k}{k!}\;,
\end{equation}
where the $(p)_k$ etc are binomial coefficients. These manipulations allow to
factorize explictly the $\alpha$ and $\beta$ integrals in the simple form:

\begin{equation}
\int_0^1 dx \: x^p (1-x)^q = 
\frac{\Gamma(1+p)\Gamma(1+q)}{\Gamma(2+p+q)} \;,
\label{intFP}
\end{equation}
with $x=\alpha,\beta$.  with various possible values of $p, q$ (which in the
framework of dimensional regularization is valid for any values of $p, q$,
since it will eventually exhibits the location of the different poles in
$\epsilon = 0$).  Next it is just a matter of systematic algebra, using for
instance Mathematica \cite{math}, to perform all these simple integrals,
re-summing the resulting power series to all orders, while picking up the
divergent and finite pieces we need.  Actually the divergent part only occurs
only in the first two terms of the power expansion in $z$ of the $_2F_1$
function, which is simple to extract analytically. In contrast, the finite
part results from contributions of all terms, which can be extracted
nuerically as the corresponding series converges fastly. Expanding all factors
contained in Eq. (\ref{3loopab}) for $\epsilon\to 0$ finally yields
\begin{equation}
\frac{\Delta V_{\rm eff}^{(c,d)}}{N} =  \delta^2 \lambda^2 
\frac{\eta^5}{30 \pi^3}
\frac{4}{N} \left ( 1- \frac{1}{4N} \right )
\left(\frac{1}{\epsilon} +\frac{41}{10} 
- 4 \ln 2 -\frac{X}{2} - 3 \ln \frac{\eta}{M} \right)\;,
\label{3loopend}
\end{equation}
where $X\sim 1.63669$ is a constant coming from the integrations. The
renormalization of the bare three-loop result, Eq. (\ref{3loopend}), is
performed in the next appendix by introducing appropriate counterterms.

\section{Renormalization of the three-loop free energy}

Here, we give details on the renormalization procedure used.  At
order-$\delta^2$ new counterterms, which do not posses the form of the
original tree-level Lagrangian, appear. This clearly is a manifestation,
within our OPT framework, of perturbative non-renormalizability. These
counterterms originate from higher momentum-dependent divergences of some of
the two-loop fermion self-energies shown in Fig. \ref{graphs1}, and then
potentially enter as contributions to the three-loop effective potential (or
equivalently the free energy). Since the model is renormalizable in the $1/N$
expansion, it is clear that these non-renormalizable counterterms are
perturbative artifacts, which can be seen indeed to disappear in the
corresponding $1/N$ quantities, as we will briefly show below.  But since we
are not using the $1/N$ expansion explicitly in our framework we have to deal
to some extent with these non-renormalizable terms.  As we shall see below,
however, the non-renormalizable contributions to the effective potential
actually vanish, such that only standard mass, wave function, and zero point
energy counterterms are necessary to cancel the divergences (though this
feature is simply an accident of the effective potential calculated at the
second perturbative order).
\begin{figure}[htb]
  \vspace{0.5cm}
  \epsfig{figure=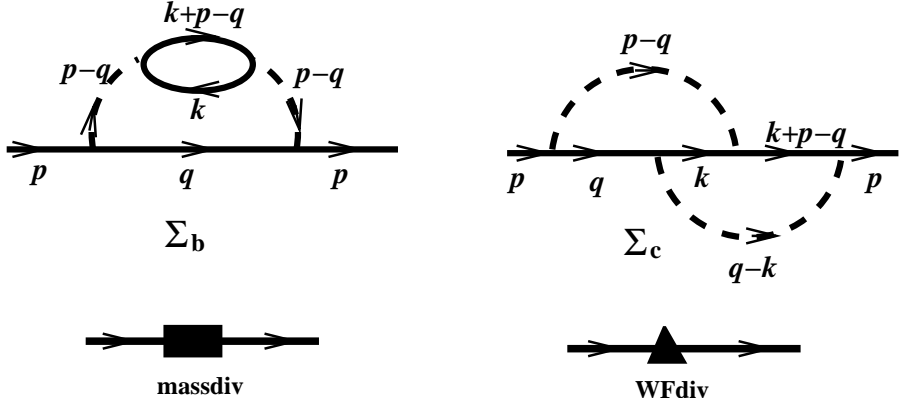,angle=0,width=12cm}
\caption[]{\label{ssuns} Momentum dependent two loop graphs contributing 
  to the fermionic self energy to order-$\delta^2$. $\Sigma_b(p)$ is of order
  $1/N$ while $\Sigma_c(p)$ contributes with $1/N^2$. The other two graphs
  represent the mass and the wave function counterterms as indicated in the
  figure.}
\end{figure}

At order-$\delta^2$ we have three two-loop contributions to the fermionic self
energy as shown if {}Fig. \ref{graphs1}.  {}From those, only the two displayed
in figure \ref{ssuns} are divergent and will be evaluated here in detail.
{}First, note that by choosing the momenta routing in an appropriate manner as
indicated in {}Fig. \ref{ssuns} one only needs to evaluate the first graph
since the closed fermionic loop contributes with $-4N$. One may explicitly
verify that $\Sigma_b(p) + \Sigma_c(p) = [1-1/(4N)] \Sigma_b(p)$.  With the
LDE {}Feynman rules at $T=0$ and $\mu=0$ one has \footnote{ The minus sign due
  to the fermionic loop has already been taken into account in Eq.
  (\ref{ssuninitial}).}

\begin{equation}
-i \Sigma_b(p)= -(-i\delta)^4\left ( \frac{-i\lambda}{\delta N} \right )^2 
\int \frac {d^d k}{(2\pi)^d} \frac {d^d q}{(2\pi)^d} 
\frac{i(\not \! q + \eta)}{(q^2 - \eta^2)}
{\rm tr}\left \{ \frac{i(\not \! k + \eta)}{(k^2 - \eta^2)}
\frac{i[(\not \! k + \not \! p -
\not \! q) + \eta]}{[(k+p-q)^2 - \eta^2]} \right \}\;.
\label{ssuninitial}
\end{equation}
Here all integrals are done in arbitrary dimension $d=3-\epsilon$.  Then,
after taking the trace we have

\begin{equation}
\Sigma_b(p) = - \delta^2 \frac{4 \lambda^2}{N}  (i)^2 \int 
\frac {d^d k}{(2\pi)^d} \frac {d^d q}{(2\pi)^d}
\frac{(\not \! q + \eta)}{(q^2 - \eta^2)}\frac{[k^2 + k(p-q) + \eta^2]}
{(k^2 - \eta^2)[(k+p-q)^2 - \eta^2]}\;.
\label{ssuntrace}
\end{equation}
Now one can introduce one Feynman parameter ($\alpha$) to merge the two $k$
dependent propagators. Then, one performs a Wick rotation (using dimensional
regularization in the $\overline {\rm MS}$ scheme) to carry on the integral
over $k$. This yields

\begin{eqnarray}
\Sigma_b(p) &=& - \delta^2 \frac{4 \lambda^2 }{N(4\pi)^{3/2}} 
(e^{\gamma_E} M^2)^{\epsilon/2}\Gamma(\epsilon/2 - 1/2)(2-\epsilon) 
\int_0^1 d\alpha
[\alpha(1-\alpha)]^{1/2-\epsilon/2} \nonumber \\
&\times&\int \frac {d^d q_E}{(2\pi)^d}\frac{-\not \! q_E + 
\eta}{(q_E^2 + \eta^2)[(p-q)_E^2 + \eta^2_\alpha]^{-1/2+\epsilon/2}}\;,
\end{eqnarray}
where $\eta^2_\alpha = \eta^2/[\alpha(1-\alpha)]$.  Now, one can use a second
{}Feynman parameter ($\beta$) to merge the two final propagators obtaining

\begin{equation}
\Sigma_b(p) = - \delta^2 \frac{4 \lambda^2 }{N(4\pi)^{3}} (e^{\gamma_E} 
M^2)^{\epsilon}\Gamma(\epsilon - 1)(2-\epsilon)\int_0^1 d\alpha
d \beta [\alpha(1-\alpha)]^{1/2-\epsilon/2}(1-\beta)^{-3/2+\epsilon/2}
\frac{[\not \! p(1-\beta) + \eta]} {\left(\eta^2_{\alpha,\beta}
\right)^{\epsilon-1}}\;,
\label{ssun1}
\end{equation}
where

\begin{equation}
\eta^2_{\alpha,\beta} = -p^2(1-\beta)\beta + \eta^2 \beta + \eta_\alpha^2 
(1-\beta)\;.
\end{equation}
Note that in the above equations we have already returned to Minkowski space.
Let us now examine the type of divergences and corresponding counterterms
emerging from these two-loop fermion self-energy expressions.  The
counterterms appear in the Lagrangian density ${\cal L}^\delta_{\rm ct}$

\begin{equation}
{\cal L}^\delta_{\rm ct} =
\bar{\psi}_{k} \left[ i \not\!\partial\:
A^\delta(\eta)\right] \psi_{k} +
B^\delta(\eta) {\bar \psi_k} \psi_k\;,
\label{GNct}
\end{equation}
which is to be added to the original Lagrangian density, Eq. (\ref {GNlde}).
The perturbative order at which the required counterterm contributions enter
first can be readily found from Eq. (4.5):

\begin{equation}
\frac{V_{{\rm eff},{\rm ct}}}{N} (\sigma_c,\eta)=
i  \int \frac {d^d p}{(2\pi)^d} {\rm tr}\ln \left[\not \! p(1+ A^\delta )
- \eta(1+B^\delta \right]\;,
\label{generalct}
\end{equation}
where, implicitly, those counterterm coefficients are of order $(\delta
\lambda)^2$ (since at lowest $\lambda\delta$ order, all calculations are
finite in dimensional regularization as we saw in section III and IV).  These
counterterms are extracted from the $\not \!p$ and $\eta$ dependent terms in
Eq. (\ref {ssun1}). {}For the $\not \!p$ term for instance:

\begin{equation}
[\Sigma_b(p)+ \Sigma_c(p)]_{\rm WF div} =
\frac{- \delta^2\lambda^2}{(4\pi)^{3}}\frac{4}{N}\left ( 1-
\frac{1}{4N} \right ) (e^{\gamma_E} M^2)^{\epsilon}
\Gamma(\epsilon - 1)(2-\epsilon)\int_0^1 d\alpha \,
d \beta \, [\alpha(1-\alpha)]^{\frac{1}{2}-\frac{\epsilon}{2}}
(1-\beta)^{-\frac{1}{2}+
\frac{\epsilon}{2}}
\frac{{\not \! p} }{\left( \eta^2_{\alpha,\beta} \right)^{\epsilon-1}}\;,
\label{ssunwfct}
\end{equation}
where the only pole is contained in $\Gamma(\epsilon-1)$.  One can then expand
in $\epsilon$ and perform the (finite) integrals is $\alpha$ and $\beta$ to
obtain a counterterm as ${\not \! p} A^{\delta}(\eta,p)$ where

\begin{equation}
A^{\delta}(\eta,p)= -\delta^2 \frac{\lambda^2}{(4\pi)^{3}}
\frac{4}{N}\left ( 1- \frac{1}{4N} \right )\frac{\pi}{15 \epsilon} 
\left( -p^2 +25\eta^2 \right) \;.
\label{wfcoef}
\end{equation}
Similarly the other counterterm can be extracted from the $\eta$ dependent
term in Eq. (\ref {ssun1}):

\begin{equation}
[\Sigma_b(p)+ \Sigma_c(p)]_{\rm mass div} =
 \frac{-\delta^2 \lambda^2}{(4\pi)^{3}}\frac{4}{N}\left ( 1
- \frac{1}{4N} \right
 ) (e^{\gamma_E} M^2)^{\epsilon}\Gamma[\epsilon - 1](2-\epsilon)
\int_0^1 d\alpha \,
d \beta \, [\alpha(1-\alpha)]^{\frac{1}{2}-\frac{\epsilon}{2}}
(1-\beta)^{-\frac{3}{2}+
\frac{\epsilon}{2}}
\frac{{\eta} }{ \left( \eta^2_{\alpha,\beta} \right)^{\epsilon-1}}\;.
\label{ssunct}
\end{equation}

The same type of manipulations that lead to the pole in Eq.  (\ref {wfcoef})
above fixes the other counterterm as $\eta B^{\delta}(\eta,p)$ where
\begin{equation}
B^{\delta}(\eta,p)= \delta^2 \frac{\lambda^2}{(4\pi)^{3}}\frac{4}{N}
\left ( 1- \frac{1}{4N} \right )\frac{\pi}{3 \epsilon} 
\left( -p^2+9\eta^2 \right)\;.
\label{etacoef}
\end{equation}
Note in Eqs. (\ref{wfcoef}) and (\ref{etacoef}) the extra $p^2$ dependence
which is the manifestation of the perturbative non-renormalizability.

\begin{figure}[htb]
  \vspace{0.5cm}
  \epsfig{figure=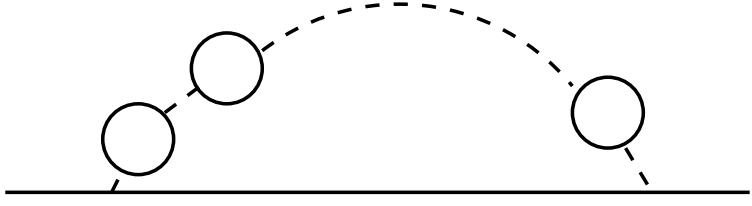,angle=0,width=10cm}
\caption[]{\label{figsumN} Next-to-leading order in the $1/N$ expansion
  graphs contributing to the fermionic self energy.}
\end{figure}
As mentioned above, all such non-renormalizable terms should be absent in the
renormalizable $1/N$ expansion.  More precisely, one can check this at the
next-to-leading $1/N$ order where the equivalent contribution to the fermion
self-energy is shown in Fig. \ref{figsumN} (which corresponds to an infinite
sum of perturbative one-loop graphs).  We can easily obtain the structure of
the divergent terms, replacing expression (\ref{wfcoef}) and (\ref{etacoef}),
in dimensional regularization as:

\begin{equation}
 \frac{1}{N\,\pi^2} \, \frac{1}{\epsilon}\; 
\left( \frac{\not \! p}{3} +\eta \right)\;,
\label{1/NCT}
\end{equation}
where we used expression 2.12b of Ref. \cite{Park} for the $d=2+1$ resummed
$\sigma$-propagator. The expression (\ref{1/NCT}) is consistent with the
calculation of wave-function and mass counterterm as performed in Ref.
\cite{Park} with cutoff regularization. The crucial point in the $1/N$
expansion is that the resummed $\sigma$ propagator has the non-perturbative
$|p|^{-1}$ behavior \cite{Park} at large $|p|$, thus damping the degree of
divergences with respect to usual perturbative graphs (such as those in Fig.
14) and thus the absence of momentum dependent divergences in Eq.
(\ref{1/NCT}).  Nevertheless, since our construction relies on standard
perturbation, in principle we would have to introduce corresponding new
counterterms, with higher derivatives, into the Lagrangian at this
perturbative order.  But actually these $p$-dependent pieces contributions are
canceled in the free energy graphs, as we will see next, so that it is not
needed to deal with these extra non-renormalizable terms. Inserting the above
two-loop counterterms into the (one-loop) free energy graph, for instance for
the $A$ counterterm in Eq.  (\ref{wfcoef}) yields

\begin{equation}
\frac{V_{\rm eff}^A}{N} = -i\int \frac {d^d p}{(2\pi)^d}{\rm tr}\frac
{{\not \! p} -\eta}{ p^2 -\eta^2  }A^\delta(\eta)\;,
\end{equation}
which gives, after taking the trace and going to Euclidean momenta:

\begin{equation}
\frac{V_{\rm eff}^A}{N} = 
\delta^2\frac{8\lambda^2\eta^2}{5 \epsilon (4\pi)^{2}}\frac{4}{N}
\left ( 1- \frac{1}{4N} \right ) \int \frac {d^dp_E}{(2\pi)^d} \frac{p_E^2}
{p_E^2 + \eta^2}\;,
\label{Ewfctint}
\end{equation}
where we simply used for the integrand $p^2/(p^2+\eta^2) = 1
-\eta^2/(p^2+\eta^2)$. Note also that the integral $\int d^n p p^2$ is zero in
dimensional regularization, so that only a mass dependent divergence remained
in Eq. (\ref{Ewfctint}) and we finally obtain

\begin{equation}
\frac{V_{\rm eff}^A}{N} = \delta^2\eta^5 
\frac{8\lambda^2}{5 (4\pi)^{3}}\frac{4}{N}
\left ( 1- \frac{1}{4N} \right ) \left [ \frac{1}{\epsilon} 
- \ln \left ( \frac{\eta}{M} \right ) +1-\ln 2 \right]\;.
\label{Ewfct}
\end{equation}
Similarly, for

\begin{equation}
\frac{V_{\rm eff}^B}{N} = i\int \frac {d^dp}{(2\pi)^d}{\rm tr}\frac
{{\not \! p} -\eta}{p^2 -\eta^2}B^\delta(\eta)
\;,
\end{equation}
one obtains \footnote{Note that the final mass and wave function counterterms
  are proportional to $\lambda^2\eta^2$, which simply reflects that the
  coupling $\lambda$ of the 3-dimensional GN model has mass dimension $-1$ ,
  i.e. the counterterms coefficients are actually dimensionless.}

\begin{equation}
\frac{V_{\rm eff}^B}{N} =- \delta^2\eta^5 \frac{8\lambda^2}{3  
(4\pi)^{3}}\frac{4}{N}\left ( 1- \frac{1}{4N} \right )
\left [ \frac{1}{\epsilon} - \ln \left ( \frac{\eta}{M} 
\right ) +1 -\ln 2 \right]\;.
\label{Emct}
\end{equation}

Next, the contributions in Eqs. (\ref{Ewfct}), (\ref{Emct}) should be added in
the $\overline{\rm MS}$ scheme to the bare three loop free energy expression
(\ref{3loopend}). This leaves a remaining divergent contribution:

\begin{equation}
\frac{V_{\rm eff}^{c+d+A+B}}{N} = \delta^2\eta^5 \frac{\lambda^2}
{15  (4\pi^{3})}\frac{4}{N}\left ( 1- \frac{1}{4N} \right ) \frac{1}{\epsilon}
\;, 
\end{equation}
which is finally renormalized by an additional vacuum (free energy)
counterterm. This additional counterterm is expected, since the free energy is
a composite operator, and is needed similarly in the (renormalizable) $d=1+1$
GN model, as shown in a previous reference \cite{prdgn2}.  Note that, although
the perturbative non-renormalizability manifests itself at order $\delta^2$ by
the presence of the higher derivative divergences in Eqs.  (\ref{wfcoef}),
(\ref{etacoef}), corresponding counterterms are not needed for the quantities
(and perturbative order) we restrict ourselves to. Thus the arbitrariness in
the final renormalized free energy is not more than the usual renormalization
scale introduced from dimensional regularization.  The complete renormalized
three-loop contribution to the free energy is obtained, in the $\overline{\rm
  MS}$ scheme, by adding all the finite contributions as given in Eqs.
(\ref{fingraph1}), (\ref{fingraph2}), (\ref{Ewfct}), (\ref{Emct}),
respectively, leads to the final expression, Eq.  (\ref{Eren3}).

\end{document}